\documentclass[aps,prb,amsmath,amssymb,superscriptaddress,twocolumn]{revtex4-2}
\usepackage{braket}
\usepackage[T1]{fontenc}
\usepackage{graphicx}
\usepackage{hyperref}
\usepackage{bbm,color,ulem}
\DeclareMathAlphabet{\mathpzc}{OT1}{pzc}{m}{it} 
\begin{document}
\title{Effects of leakage on the realization of a discrete time crystal in a chain of singlet-triplet qubits}
\author{Robert E.\ Throckmorton}
\author{S.\ \surname{Das Sarma}}
\affiliation{Condensed Matter Theory Center and Joint Quantum Institute, Department of Physics, University of Maryland, College Park, Maryland 20742-4111 USA}
\date{\today}
\begin{abstract}
We consider the effects of leakage on the ability to realize a discrete time crystal (DTC) in a semiconductor quantum dot linear array being operated as a chain of singlet-triplet (ST) qubits.  This system realizes an Ising model with an effective applied magnetic field, plus additional terms that can cause leakage out of the computational subspace.  We demonstrate that, in the absence of these leakage terms, this model theoretically realizes a DTC phase over a broad parameter regime for six and eight qubits, with a broader parameter range for the eight-qubit case.  We then reintroduce the leakage terms and find that the DTC phase disappears entirely over the same parameter range if the system is only subject to a uniform magnetic field, which does not suppress leakage.  However, we find that the DTC phase can be restored if the system is instead subject to a magnetic field that alternates from qubit to qubit, which suppresses leakage.  We thus show that leakage is a serious problem for the realization of a DTC phase in a chain of ST qubits, but is by no means insurmountable.  Our work suggests that experiments manifesting small-system stable DTC should be feasible with currently existing quantum dot spin qubits.
\end{abstract}
\maketitle

\section{Introduction}
Even though the system size and gate fidelities in semiconductor-based spin qubits lag behind those of other platforms (e.g., superconducting or ion trap) for quantum computation, much progress has been made in improving the fidelities of single- and two-qubit gates, with experiments coming close to, or even exceeding, the $99.9\%$ threshold needed to implement some error-correcting techniques \cite{nnano.2014.216,nature15263,science.aao5965,s41928-019-0234-1,s41586-019-1197-0.2019,acs.nanolett.0c02397,acs.nanolett.0c04771,nature25766,PhysRevX.9.021011,arxiv.2111.11937,s41586-021-04182-y,s41586-021-04273-w,arxiv.2202.09252}, which are essential to building a functioning quantum computer.  The systems used in these experiments, however, are small, with the largest only having six qubits \cite{arxiv.2202.09252}.  One may ask what else can be done with these small systems besides quantum computation experiments.  One possibility is to realize a discrete time crystal (DTC) with them, which is the application that we will focus on.

In recent years, DTCs have been a topic of great interest, both theoretically and experimentally.  The general concept of a time crystal was first proposed in 2012 by Wilczek \cite{PhysRevLett.109.160401,PhysRevLett.109.160402} as an analog to a conventional (space) crystal; just as crystals break continuous space translation symmetry, a time crystal breaks continuous time translation symmetry.  It would later be proved that the spontaneous breaking of continuous time translation symmetry needed to realize a time crystal in thermal equilibrium, as envisioned by Wilczek, is impossible in a large class of systems because of a no-go theorem explicitly ruling out the breaking of the continuous time translation symmetry\cite{10.1209/0295-5075/103/57008,PhysRevLett.111.070402,PhysRevLett.114.251603}.  However, it should be possible, under well-defined conditions, to spontaneously break a {\it discrete} time translation symmetry, found in periodically driven systems, leading to a DTC phase.  A number of theoretical works have already investigated the existence of DTC phases in such periodically driven systems \cite{PhysRevB.99.035311,PhysRevB.101.115303,arxiv.2111.14327,arxiv.2209.05510,Sarkar2022,acs.nanolett.2c00976}.  In addition, a number of experiments have found evidence for DTCs in periodically-driven qubit systems \cite{nature21413,s41586-021-04257-w,science.abk0603,arxiv.2108.00942}, and another experiment reported a time crystal state in a Bose-Einstein condensate \cite{science.abo3382}.  In the current work, we do not ask whether a stable and robust time crystal can exist for infinite time in the thermodynamic limit (it now appears that most likely DTC is a long-lasting transient rather than a thermodynamic phase), which is an important question of principle, but focus on the possible laboratory realization of DTC in small systems (of semiconductor quantum dots) for reasonably long times of experimental relevance.

A DTC phase is defined by two properties that must be satisfied for any initial condition.  First, the Hamiltonian must be periodic with period $T$, but the system's response must not itself be periodic with the same period, i.e., $H(t)=H(t+T)$, but $\ket{\psi(t)}\neq\ket{\psi(t+T)}$.  Instead, $\ket{\psi(t)}=\ket{\psi(t+nT)}$, where $n$ is an integer and $n>1$.  Typically, $n=2$, so that we observe period doubling.  Second, and most importantly, the period of the response must be robust against imperfections in the drive (e.g., fluctuations in amplitude or timing of the drive).  This, of course, parallels the rigidity of a space crystal; just as a small perturbation to the position of an atom in a crystal will not destroy the crystalline structure, so a small disturbance to the perfect periodicity of a drive should not eliminate the periodic response.  In the context of a qubit system, the (ideal) drive is a perfectly periodic sequence of pulses that implement $\pi$ rotations of all of the qubits, so an imperfection could include a mistiming of a pulse or a pulse that instead implements a $(1-\epsilon)\pi$ rotation.  A number of criteria have been identified for qubit systems that exhibit a DTC phase \cite{PRXQuantum.2.030346}, which we summarize here.  The system must exhibit many-body localized (MBL) behavior, have a long coherence time, have short-ranged interactions that are predominantly Ising in nature, and be Ising-even, i.e., the interaction terms must be of the form, $\sum_{ij}J_{ij}Z_iZ_j$, and the full Hamiltonian must commute with the Ising operator, $\prod_{i}X_i$.

We consider here a chain of Heisenberg exchange-coupled spins in semiconductor quantum dots subject to an applied magnetic field, with the exchange couplings and magnetic field arranged so that the system operates as a chain of singlet-triplet (ST) qubits \cite{PhysRevLett.89.147902,science.1116955}.  We assume the presence of quasistatic noise in the exchange couplings and magnetic field gradients, modeled here as Gaussian distributions.  We set the interqubit exchange couplings to be much larger than the intraqubit couplings, and we set the intended magnetic field gradients to zero (i.e., field gradients only occur because of noise).  If we rewrite the Hamiltonian in terms of the computational states, $\ket{0}=\tfrac{1}{\sqrt{2}}(\ket{\uparrow\downarrow}-\ket{\downarrow\uparrow})$ and $\ket{1}=\tfrac{1}{\sqrt{2}}(\ket{\uparrow\downarrow}+\ket{\downarrow\uparrow})$, and the leakage states, $\ket{L_+}=\ket{\uparrow\uparrow}$ and $\ket{L_-}=\ket{\downarrow\downarrow}$, of the qubits, then we find that the system would realize an Ising model if not for the leakage terms.  Such a model would be ideal for realizing a DTC, as it meets the criteria listed above: we obviously have an Ising interaction and can tune the parameters to make the Hamiltonian (at least approximately) Ising-even, and the exchange interactions are short-ranged.  We will see that the Ising interaction has the form, $-\sum_{ij}J_{ij}X_iX_j$, which is just the form given earlier in a rotated basis; in this case, the Ising operator is instead $\prod_{i}Z_i$.  We are thus interested in what the effects of these leakage terms are and whether or not it is possible to mitigate their detrimental effects.

To this end, we consider the periodic application of pulses, with period $T$, to this system that implement a $(1-\epsilon)\pi$ rotation on all qubits (we thus have included in error term $\epsilon$ in what would ideally be a $\pi$ rotation).  We consider systems consisting of six and eight qubits, and consider four different initial conditions for each system size.  We fix the strength of the noise in the magnetic field gradients $\sigma_{\delta h}$, defined here as the standard deviation of the Gaussian distribution, and vary the strength of the noise in the interqubit exchange couplings $\sigma_{J'}$ and $\epsilon$.  We calculate the Bloch sphere positions of all qubits as a function of the number of Floquet periods.  We then determine which initial conditions, if any, exhibit DTC behavior.  We make this determination by finding the Fourier transforms of the Bloch sphere positions of all of the qubits and looking for a peak corresponding to oscillations of period $2T$.  If all qubits have this peak for all four initial conditions, then we declare the system to be in a DTC phase.  If only some of the initial conditions show such behavior in all qubits, then we consider the system to be in a ``pre-thermal'' phase.  Finally, if none of the intial conditions exhibit DTC behavior, then we conclude that the system is in a ``thermal'' phase with no DTC at all.

We first consider the limit in which the leakage terms are dropped, yielding a pure Ising model in an (effective) applied magnetic field.  This is the ideal starting point, which is necessary for benchmarking the realistic experimentally relevant situations.  In this case, we find a DTC phase over a large parameter range for both six and eight qubits.  Increasing the number of qubits expands the parameter range over which the DTC phase appears.  We then introduce the leakage terms.  We find that the DTC phase disappears completely if the system is subject only to a uniform magnetic field, thus showing that leakage is a serious problem for the realization of a DTC in a chain of ST qubits.  In general, ST qubit systems would not manifest DTC although the Hamiltonian is mostly Ising-like.  However, we show that it is possible to restore the DTC phase by applying a strong alternating magnetic field to each qubit; if the applied alternating magnetic field has an energy scale much larger than the interqubit exchange coupling, then the phase diagram that we obtain is hardly distinguishable from that found for the ideal no-leakage limit.  This happens because the alternating magnetic field helps to freeze out the leakage states.  We also consider other magnetic field configurations, including ``two up, two down'' (i.e., apply a magnetic field $+B$ to the leftmost two qubits, then $-B$ to the next two qubits, and so on), ``three up, three down'' (analogous to ``two up, two down''), and (for eight qubits specifically) ``four up, four down.''  In these cases, we also find suppression of the DTC phase.  We therefore believe that DTC should be realizable in quantum-dot-based spin qubits.

The rest of the paper is organized as follows.  We introduce our model in detail in Sec.~\ref{sec:Model}.  We then look for DTC states in this model both with and without the leakage terms present in Sec.~\ref{sec:DTCStates}.  Finally, we present our conclusions in Sec.~\ref{sec:Conclusion}.

\section{Model} \label{sec:Model}
The underlying physical model that we employ is that of a chain of electron spins with nearest-neighbor Heisenberg exchange couplings and applied magnetic fields:
\begin{equation}
H=\sum_{i=1}^{L-1}J_i\vec{\sigma}_i\cdot\vec{\sigma}_{i+1}+\sum_{i=1}^{L} h_i\sigma_{i,z},
\end{equation}
where $J_i$ is the exchange coupling between spins $i$ and $i+1$, and $h_i$ is the Zeeman energy of spin $i$ in the presence of an applied magnetic field.  We arrange the values of these terms to realize a chain of coupled singlet-triplet qubits as follows.  Let $N$ be the number of qubits that we want to realize, so that there are $L=2N$ spins.  We let spins $1$ and $2$ form one qubit, $3$ and $4$ the next qubit, and so on.  We set the exchange coupling between the two spins in qubit $j$ (i.e., spins $2j-1$ and $2j$) to a value $J_j$, with $j=\lceil\tfrac{i}{2}\rceil$, where $\lceil\cdot\rceil$ is the ceiling function, and the exchange coupling between the second spin of qubit $j$ and the first of qubit $j+1$ (i.e., spins $2j$ and $2j+1$) to $J'_j$.  We assume a ``staggered'' magnetic field; i.e., the magnetic field experienced by spin $i$ is given by $h_i=B_j+\tfrac{1}{2}(-1)^i\delta h_j$, so that there is a magnetic field gradient $\delta h_j$ on qubit $j$.

We now rewrite our Hamiltonian in terms of the computational and leakage states of the singlet-triplet qubits.  The computational states are $\ket{0}=\ket{S}=\tfrac{1}{\sqrt{2}}(\ket{\uparrow\downarrow}-\ket{\downarrow\uparrow})$ and $\ket{1}=\ket{T_0}=\tfrac{1}{\sqrt{2}}(\ket{\uparrow\downarrow}+\ket{\downarrow\uparrow})$, and the leakage states are $\ket{L_+}=\ket{T_+}=\ket{\uparrow\uparrow}$ and $\ket{L_-}=\ket{T_-}=\ket{\downarrow\downarrow}$.  We obtain $H=H_q+H_\text{int}+H_L$, where
\begin{equation}
H_q=\sum_{j=1}^{N}(2J_j Z_j-\delta h_j X_j+J_j1_{2,j}+2B_j\tilde{Z}_j) \label{eq:Hamiltonian_SQ_Full}
\end{equation}
are the single-qubit terms (note that, here and in the next equation, $j$ runs over qubits rather than spins),
\begin{equation}
H_\text{int}=\sum_{j=1}^{N-1}J'_j(-X_jX_{j+1}+\tilde{Z}_jX_{j+1}-X_j\tilde{Z}_{j+1}+\tilde{Z}_j\tilde{Z}_{j+1})
\end{equation}
are the interaction terms, and $H_L$, which will be fully defined shortly, are the leakage terms.  Here, $X_j$, $Y_j$, and $Z_j$ are the Pauli operators acting within the computational subspace, while $\tilde{X}_j$, $\tilde{Y}_j$, and $\tilde{Z}_j$ are those acting within the leakage subspace.

We now define the leakage terms via their action on all possible states of two neighboring qubits $j$ and $j+1$.  Their action in the case where both qubits are in computational states is
\begin{eqnarray}
&&H_L\ket{s}_j\ket{s'}_{j+1}\cr
&&=J'_j[\ket{L_+}_j\ket{L_-}_{j+1}+(-1)^{s+s'}\ket{L_-}_j\ket{L_+}_{j+1}];
\end{eqnarray}
in the case where one of the qubits is in a leakage state it is
\begin{eqnarray}
&&H_L\ket{s}_j\ket{L_{S'}}_{j+1}\cr
&&=-(-1)^{s(1-s')}2J'_j\ket{L_{S'}}_j\frac{1}{\sqrt{2}}[\ket{0}-(-1)^s\ket{1}]_{j+1}, \label{eq:MixedState1} \\
&&H_L\ket{L_S}_j\ket{s'}_{j+1}\cr
&&=-(-1)^{ss'}2J'_j\frac{1}{\sqrt{2}}[\ket{0}+(-1)^{s'}\ket{1}]_j\ket{L_S}_{j+1}, \label{eq:MixedState2}
\end{eqnarray}
where $S'$ ($S$) in Eq.~\eqref{eq:MixedState1} [Eq.~\eqref{eq:MixedState2}] is $-$ for $s'$ ($s$) equal to $0$ and $+$ if it is $1$; and in the case where both qubits are in leakage states it is
\begin{eqnarray}
H_L\ket{L_+}_j\ket{L_+}_{j+1}&=&H_L\ket{L_-}_j\ket{L_-}_{j+1}=0, \\
H_L\ket{L_+}_j\ket{L_-}_{j+1}&=&2J'_j\frac{1}{\sqrt{2}}[\ket{0}+\ket{1}]_j\frac{1}{\sqrt{2}}[\ket{0}+\ket{1}]_{j+1}, \nonumber \\ \\
H_L\ket{L_-}_j\ket{L_+}_{j+1}&=&2J'_j\frac{1}{\sqrt{2}}[\ket{0}-\ket{1}]_j\frac{1}{\sqrt{2}}[\ket{0}-\ket{1}]_{j+1}. \nonumber \\
\end{eqnarray}
Note that we allow the exchange couplings and magnetic Zeeman terms to vary from qubit to qubit.  This is because we introduce quasistatic noise in all of these parameters, which is mathematically equivalent to disorder.  We sample the values of the $J_j$, $J'_j$, and $\delta h_j$ from Gaussian distributions:
\begin{eqnarray}
f_J(J)&\propto& e^{-(J-J_0)^2/2\sigma_J^2}, J_j\in[0,\infty), \\
f_{j'}(J')&\propto& e^{-(J'-J'_0)^2/2\sigma_{J'}^2}, J'_j\in[0,\infty), \\
f_{\delta h}(\delta h)&\propto& e^{-(\delta h)^2/2\sigma_{\delta h}^2}.
\end{eqnarray}
We truncate the distributions for the exchange couplings to positive values because, for experimentally realistic situations, only positive values can be realized.  However, we do not expect that allowing negative values of the exchange couplings would alter our results.

\section{Discrete time crystal (DTC) states} \label{sec:DTCStates}
We now look for DTC states in the model just described, both in the absence of leakage terms and in their presence.  We will be considering systems of six and eight qubits and four different initial conditions for each.  If we let $\ket{\pm x}=\tfrac{1}{\sqrt{2}}(\ket{0}\pm\ket{1})$, these four conditions are, for six qubits,
\begin{eqnarray}
	\ket{\psi_{0,1}}&=&\ket{+x}_1\ket{+x}_2\ket{+x}_3\ket{+x}_4\ket{+x}_5\ket{+x}_6, \label{Eq:InitCond6_1} \\
	\ket{\psi_{0,2}}&=&\ket{+x}_1\ket{-x}_2\ket{+x}_3\ket{-x}_4\ket{+x}_5\ket{-x}_6, \\
	\ket{\psi_{0,3}}&=&\ket{+x}_1\ket{+x}_2\ket{+x}_3\ket{-x}_4\ket{-x}_5\ket{-x}_6, \\
	\ket{\psi_{0,4}}&=&\ket{+x}_1\ket{-x}_2\ket{-x}_3\ket{-x}_4\ket{-x}_5\ket{+x}_6, \label{Eq:InitCond6_4}
\end{eqnarray}
and, for eight qubits,
\begin{eqnarray}
	\ket{\psi_{0,1}}&=&\ket{+x}_1\ket{+x}_2\ket{+x}_3\ket{+x}_4\ket{+x}_5\ket{+x}_6\ket{+x}_7\ket{+x}_8, \nonumber \\ \label{Eq:InitCond8_1} \\
	\ket{\psi_{0,2}}&=&\ket{+x}_1\ket{-x}_2\ket{+x}_3\ket{-x}_4\ket{+x}_5\ket{-x}_6\ket{+x}_7\ket{-x}_8, \nonumber \\ \\
	\ket{\psi_{0,3}}&=&\ket{+x}_1\ket{+x}_2\ket{+x}_3\ket{+x}_4\ket{-x}_5\ket{-x}_6\ket{-x}_7\ket{-x}_8, \nonumber \\ \\
	\ket{\psi_{0,4}}&=&\ket{+x}_1\ket{+x}_2\ket{-x}_3\ket{-x}_4\ket{-x}_5\ket{-x}_6\ket{+x}_7\ket{+x}_8. \nonumber \\ \label{Eq:InitCond8_4}
\end{eqnarray}
For our numerical calculations, we fix $\sigma_{\delta h}=0.01J'_0$, $J_0=0.01J'_0$, and $\sigma_J=0.01\sigma_{J'}$.  We then vary $\sigma_{J'}$ from $10^{-2}J'_0$ to $10^{-0.1}J'_0$ and $\epsilon$ from $0$ to $0.26$.  We use $5,040$ realizations of noise, steps of $0.1$ for $\log_{10}\left (\sigma_{J'}/J'_0\right )$, and steps of $0.02$ for $\epsilon$ for the six-qubit case and the eight-qubit case without leakage, while we use $160$ realizations, steps of $0.2$ for $\log_{10}\left (\sigma_{J'}/J'_0\right )$, and steps of $0.04$ for $\epsilon$ in the eight-qubit case with leakage.

We then determine whether or not our system is in a DTC phase for given values of $\sigma_{J'}$ and $\epsilon$ in the following way.  For each of the above initial conditions, we let the system evolve under its Hamiltonian for a time $T$, and then apply a $(1-\epsilon)\pi$ rotation.  Here, $\epsilon$ represents an error in the qubit rotation; the ideal case, given by $\epsilon=0$, is a $\pi$ rotation of all qubits.  We perform $100$ of these Floquet cycles (evolution under the Hamiltonian followed by rotation of all qubits) and determine the components of the qubits' states on the Bloch sphere, denoted here as $P_x$, $P_y$, and $P_z$, as a function of the number of cycles.  We then determine the Fourier transforms of these components.  The signature of DTC behavior for a given initial condition that we look for is a peak at $\omega=\pi/T$ in $|P_x(\omega)|$ for all qubits, corresponding to oscillations of period $2T$.  We require that the system display DTC behavior for $\epsilon\neq 0$ and for all four of the above initial conditions in order to demonstrate robustness of the $2T$-periodic oscillations against errors in the rotations and thus to declare the system to be in a DTC phase.


\subsection{No-leakage limit}
We will first consider the no-leakage limit, in which we drop all leakage terms and terms that act on leakage states, so that the Hamiltonian becomes
\begin{equation}
H=\sum_{j=1}^{N}(2J_j Z_j-\delta h_j X_j+J_j1_{2,j})-\sum_{j=1}^{N-1}J'_jX_jX_{j+1}. \label{Eq:Hamiltonian_NL}
\end{equation}
We note that this is just the Ising model in the presence of an (effective) magnetic field.  We thus see that, if not for the leakage terms, a chain of singlet-triplet qubits would be a perfect system for realizing the Ising model and thus a DTC phase.  This Hamiltonian satisfies all of the conditions listed earlier for finding a DTC phase.  Note that the Ising interation is of the form, $-\sum_{ij} J_{ij} X_iX_j$, rather than $\sum_{ij} J_{ij} Z_iZ_j$, so that now the Ising operator is $\prod_i Z_i$.  This operator commutes with all of the terms in this Hamiltonian except for $-\sum_j\delta h_jX_j$; it is for this reason that set the intended magnetic field gradients to zero, so that any such gradient that appears in the system is due to noise.

\begin{figure*}[htb]
	\centering
	\includegraphics[width=\linewidth]{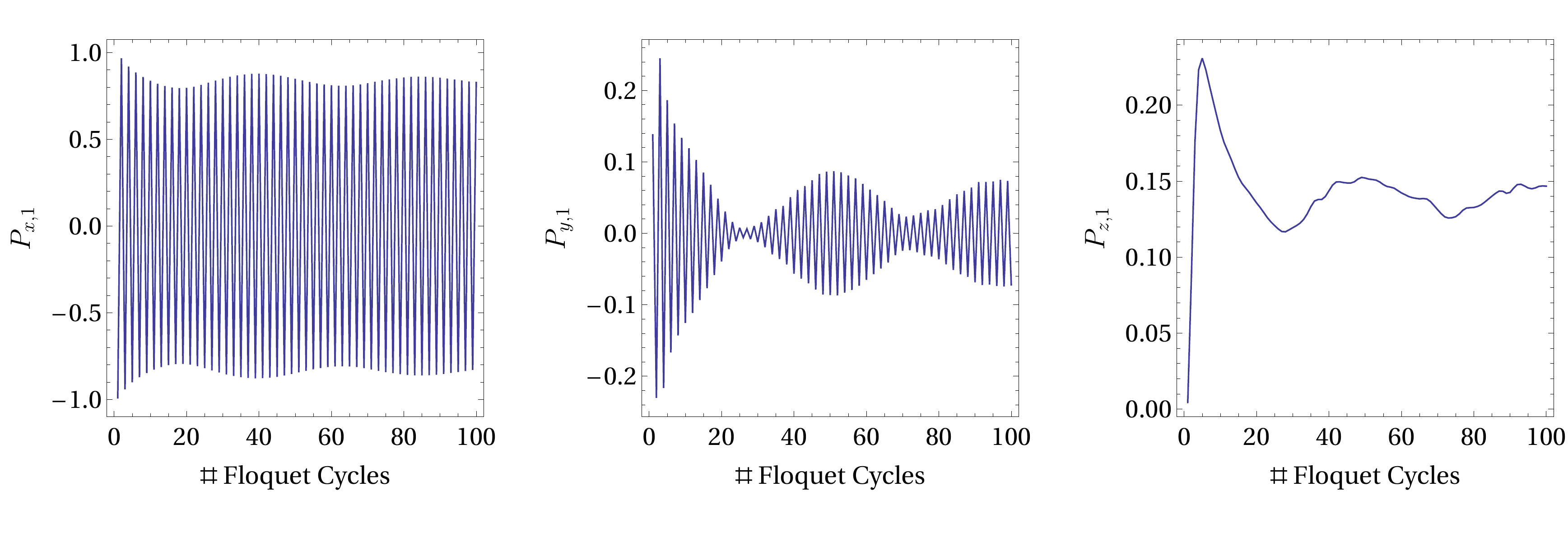}
	\includegraphics[width=\linewidth]{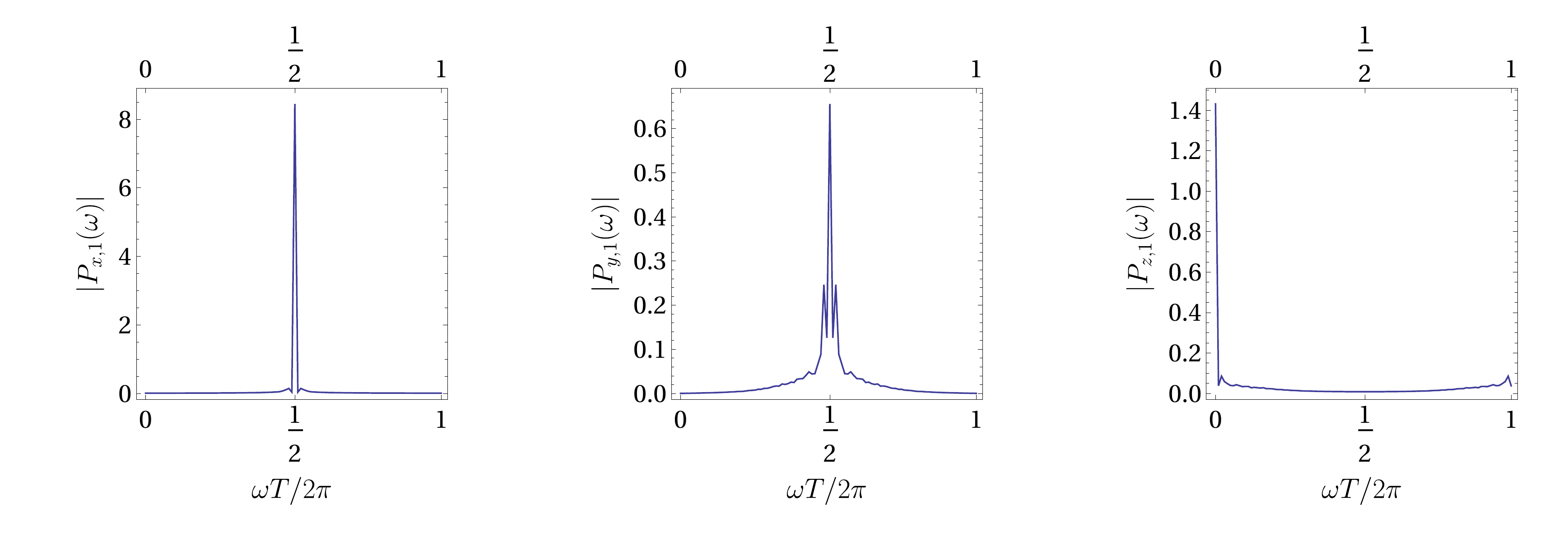}
	\caption{(Top) Example plot of the three components of the qubit state $P_x$, $P_y$, and $P_z$ for the leftmost qubit, $\sigma_{J'}=10^{-0.1}J'_0$, and for $\epsilon=0.04$. (Bottom) Absolute values of the Fourier transforms of these components.}
	\label{fig:PSPlotsExamples}
\end{figure*}

\begin{figure}[htb]
	\centering
		\includegraphics[width=\columnwidth]{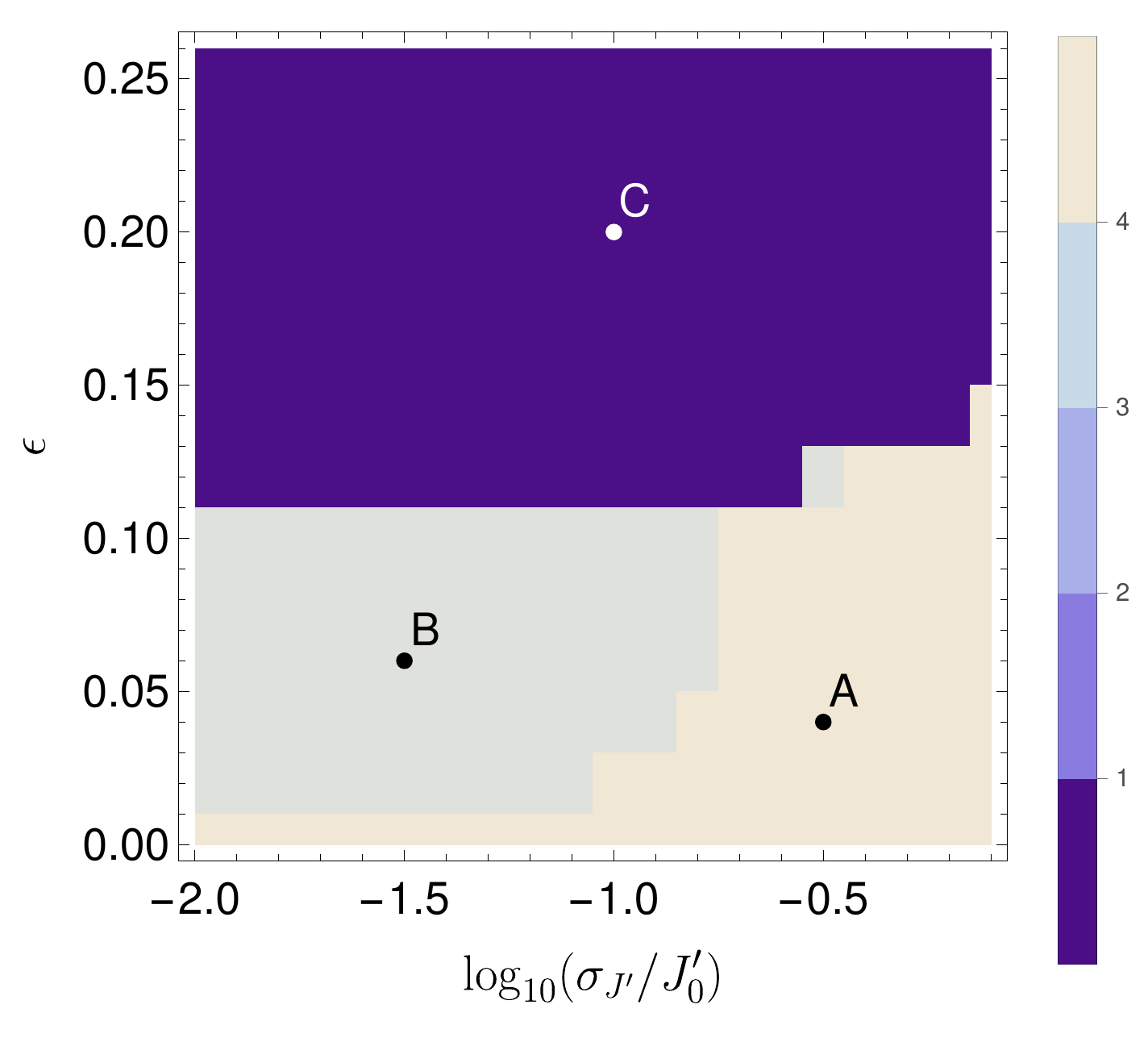}
	\caption{Plot of the number of states that display discrete time crystal (DTC) behavior as a function of $\epsilon$ and $\sigma_{J'}$ for six qubits in the no-leakage limit.  The three points labeled on this plot correspond to a DTC phase (A), a ``pre-thermal'' phase (B), and a ``thermal'' phase (C).}
	\label{fig:PDFullPlotNL_6Qubits_NoLeakage}
\end{figure}

We show plots of the three components of the Bloch sphere position of one of the qubits along with the absolute values of their Fourier transforms in Fig.~\ref{fig:PSPlotsExamples} to illustrate the peak at $\omega=\pi/T$, and then plot the phase diagram for six qubits as a function of $\sigma_{J'}$ and $\epsilon$ in Fig.~\ref{fig:PDFullPlotNL_6Qubits_NoLeakage}.

\begin{figure*}[htb]
	\centering
	\includegraphics[width=\linewidth]{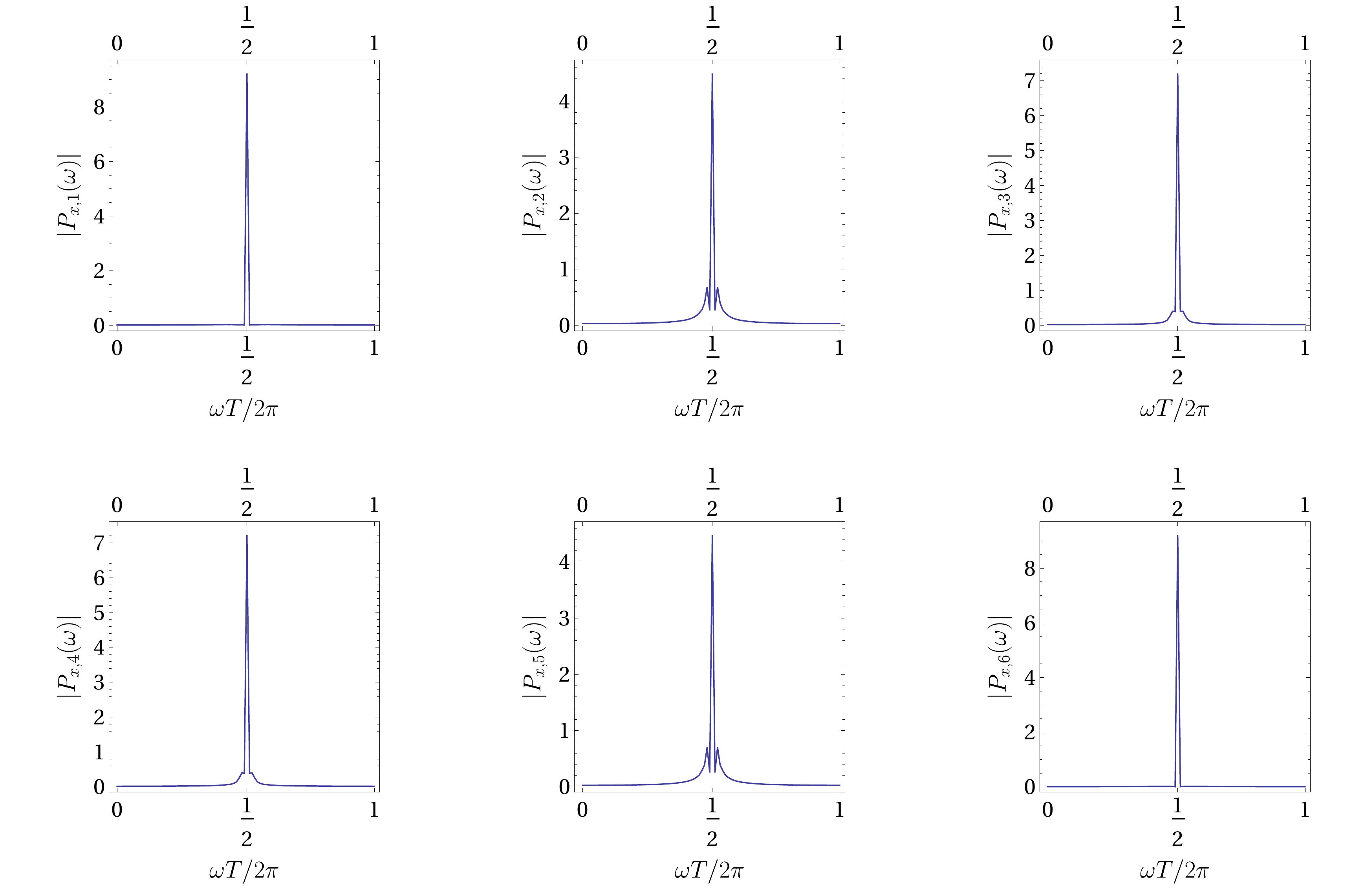}
	\caption{Plot of the Fourier transforms of $P_x$ for all six qubits for point A (DTC phase) in Fig.~\ref{fig:PDFullPlotNL_6Qubits_NoLeakage} and for the initial condition $\ket{\psi_{0,4}}$.  The plots for the other three initial conditions are similar to these, also showing peaks at $\omega=\pi/T$ for all six qubits.}
	\label{fig:PXPlots_DTCExample}
\end{figure*}

\begin{figure*}[htb]
\centering
\includegraphics[width=\linewidth]{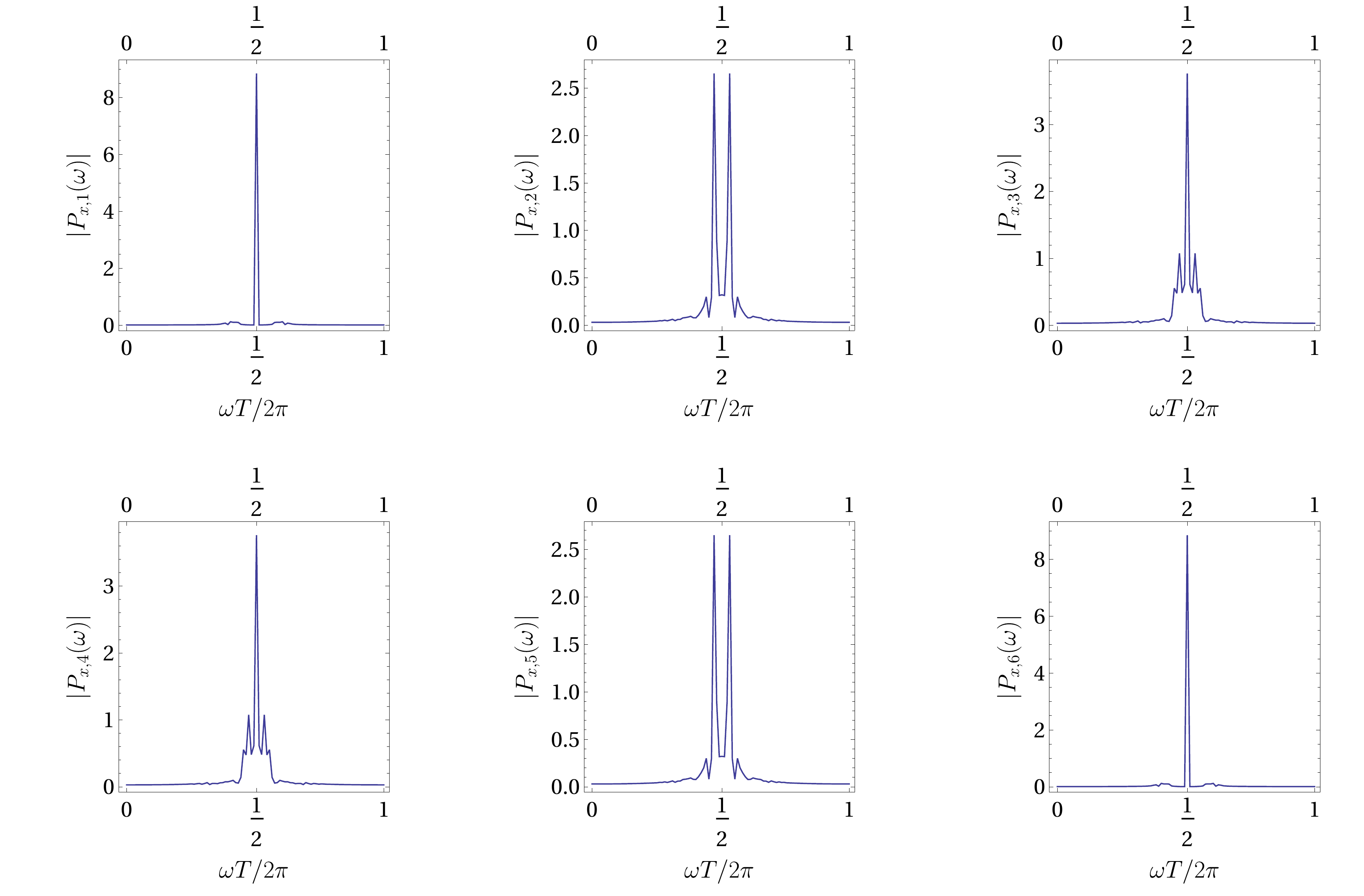}
\caption{Plot of the Fourier transforms of $P_x$ for all six qubits for point B (``pre-thermal'' phase) in Fig.~\ref{fig:PDFullPlotNL_6Qubits_NoLeakage} and for the initial condition $\ket{\psi_{0,4}}$.  Note the split peaks for qubits $2$ and $5$ for this initial condition.  The other three initial conditions yield similar results as for point A, i.e., they show DTC-like behavior.}
\label{fig:PXPlots_PreThermExample}
\end{figure*}

\begin{figure*}[htb]
\centering
\includegraphics[width=\linewidth]{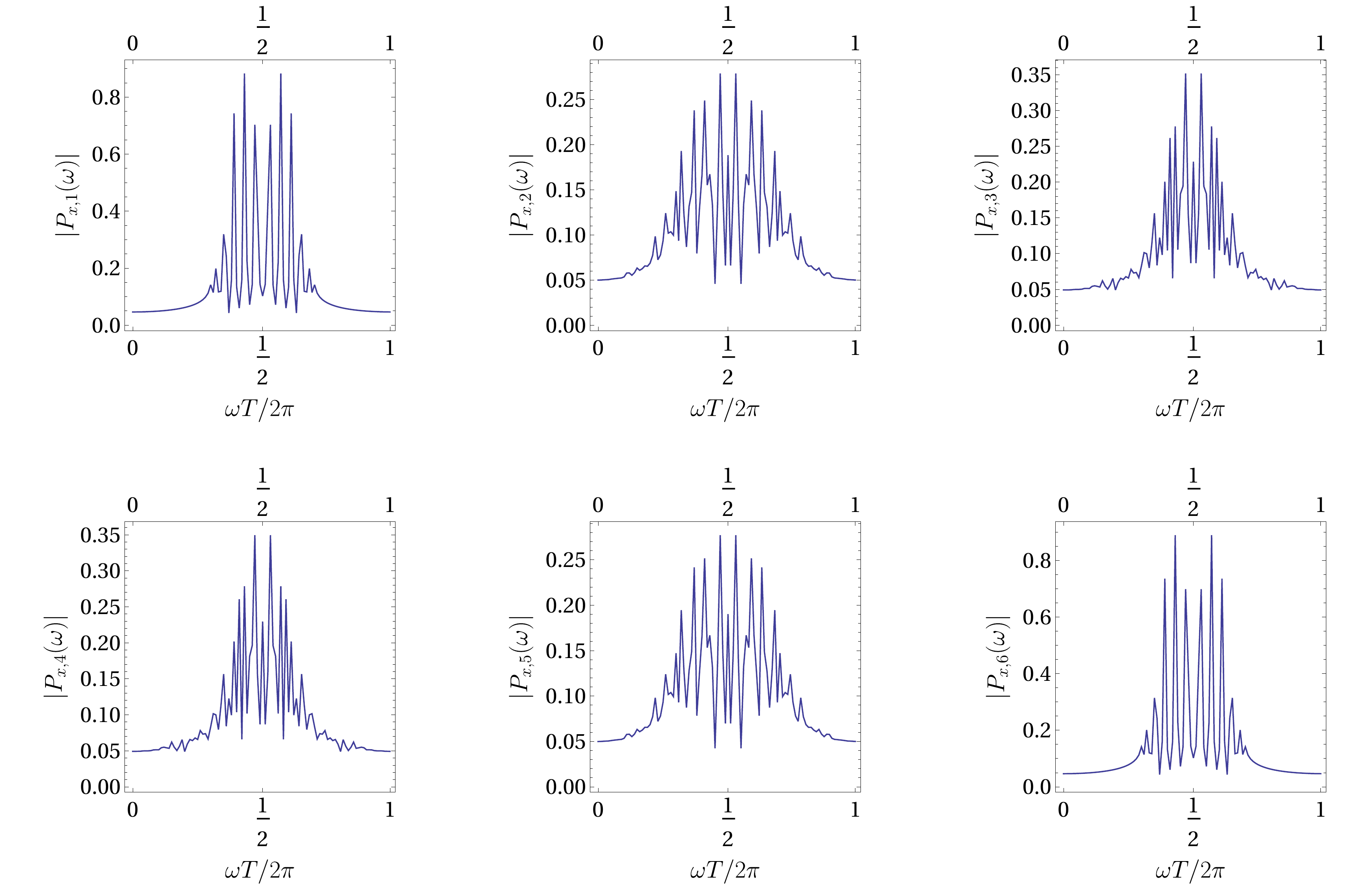}
\caption{Plot of the Fourier transforms of $P_x$ for all six qubits for point C (``thermal'' phase) in Fig.~\ref{fig:PDFullPlotNL_6Qubits_NoLeakage} and for the initial condition $\ket{\psi_{0,4}}$.  The plots for the other three initial conditions are similar to these.}
\label{fig:PXPlots_ThermExample}
\end{figure*}

We note that this phase diagram has three regions: a DTC region, a ``pre-thermal'' region in which only some of the initial conditions yield DTC-like behavior, and a ``thermal'' region in which none of the initial conditions display such behavior.  We label three points, A, B, and C, that respectively represent each of these three regions.  We show plots of $|P_x(\omega)|$ for all qubits for one initial condition, $\ket{\psi_{0,4}}$, and for six qubits, in Figs.~\ref{fig:PXPlots_DTCExample}--\ref{fig:PXPlots_ThermExample} to show examples of how the qubits behave in each of these three regions.  We use this initial condition as an example in particular because it is the one initial condition that fails to show DTC-like behavior at point B (the ``pre-thermal'' region) in Fig.~\ref{fig:PDFullPlotNL_6Qubits_NoLeakage}.

\begin{figure*}[htb]
	\centering
	\includegraphics[width=\linewidth]{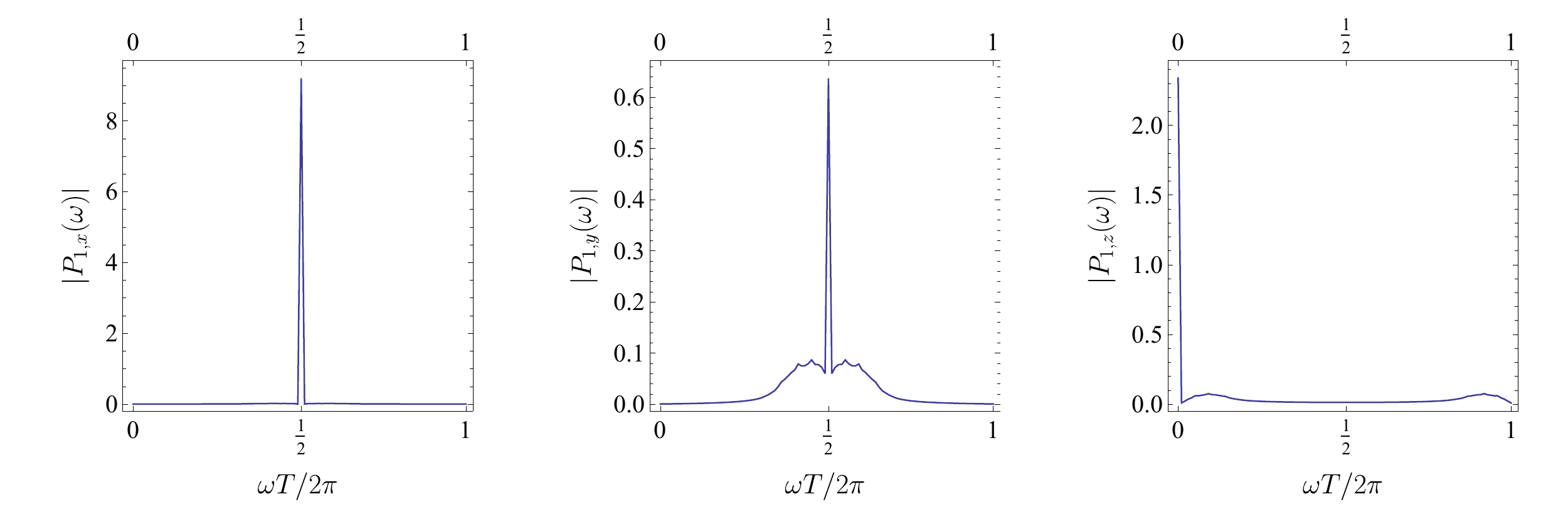}
	\includegraphics[width=\linewidth]{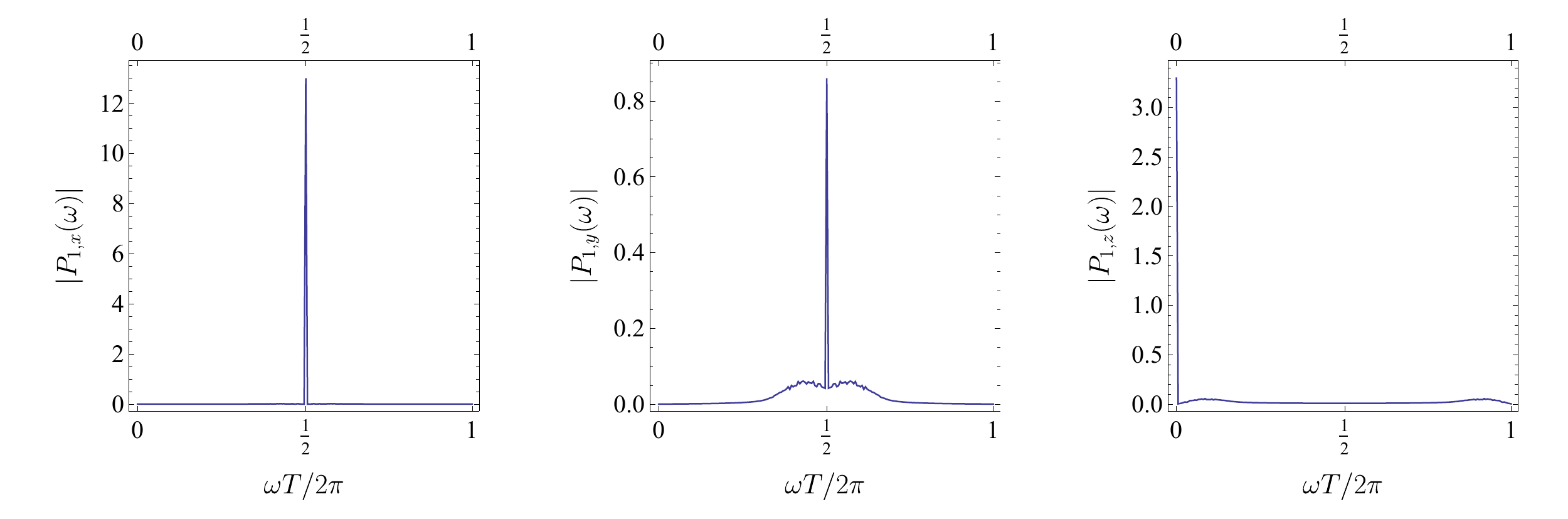}
	\caption{Plot of the Fourier transforms of $P_x$ for all six qubits for point C (``thermal'' phase) in Fig.~\ref{fig:PDFullPlotNL_6Qubits_NoLeakage} and for the initial condition $\ket{\psi_{0,4}}$.  The plots for the other three initial conditions are similar to these.}
	\label{fig:Example_ShortVsLongTime}
\end{figure*}

\begin{figure}[htb]
	\centering
	\includegraphics[width=\columnwidth]{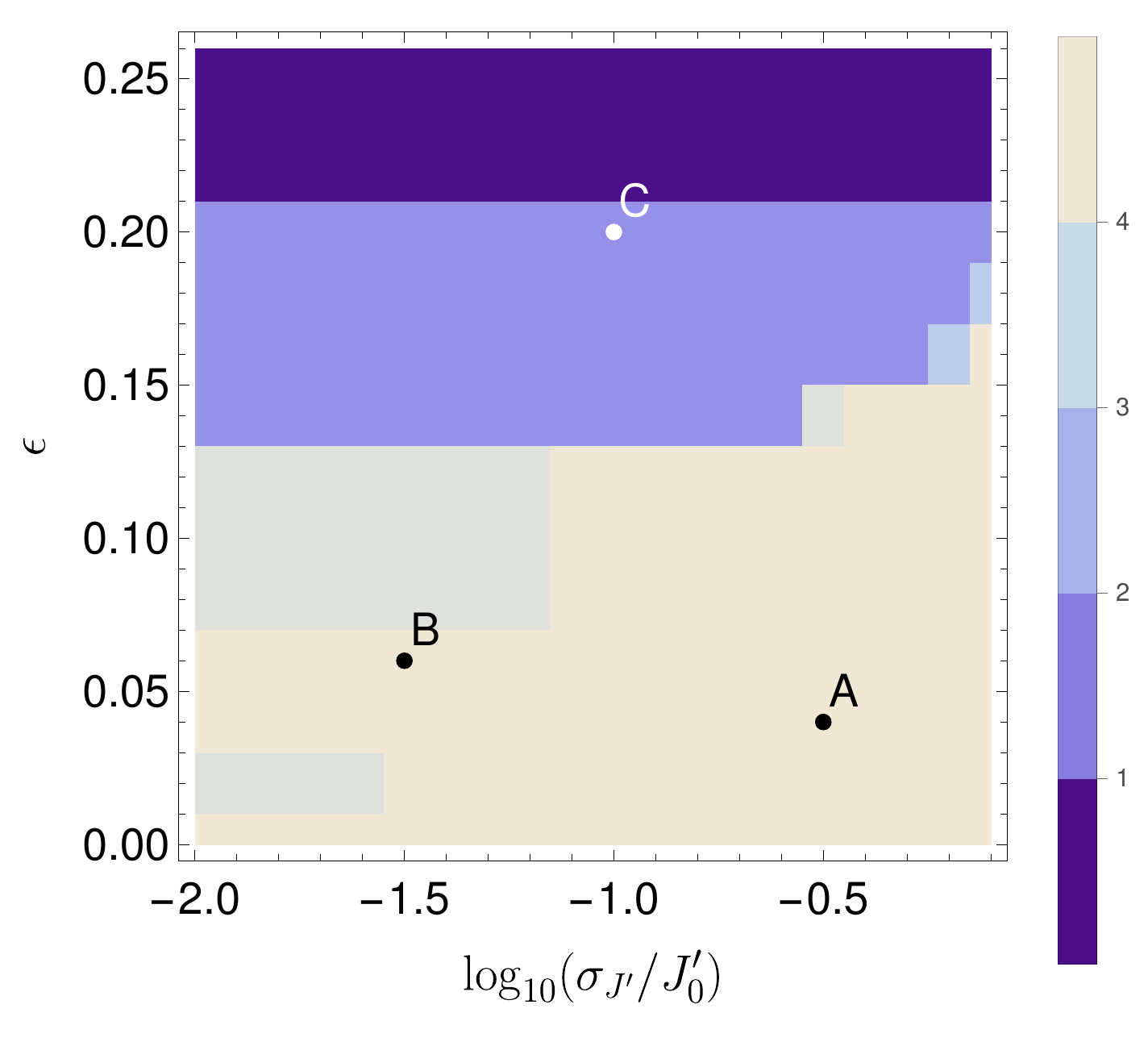}
	\caption{Plot of the number of states that display discrete time crystal (DTC) behavior as a function of $\epsilon$ and $\sigma_{J'}$ for eight qubits in the no-leakage limit.  We include the same three points, A, B, and C, as we do in the corresponding result for six qubits shown in Fig.~\ref{fig:PDFullPlotNL_6Qubits_NoLeakage}.}
	\label{fig:PDFullPlotNL_8Qubits_NoLeakage}
\end{figure}

We also investigate the effects of the number of Floquet cycles and of system size on the presence or absence of a DTC phase.  We show a comparison between our results for $100$ Floquet cycles and the corresponding results for $200$ cycles in Fig.~\ref{fig:Example_ShortVsLongTime}.  We see that the results have not changed qualitatively; the only difference is that the peaks are sharper, as expected for a larger number of Floquet cycles.  The phase diagram is also unchanged if we use $200$ cycles instead of $100$.
On the other hand, a larger system size does have a significant effect on the phase diagram; we show the results for eight qubits in Fig.~\ref{fig:PDFullPlotNL_8Qubits_NoLeakage}.  We see that, for the larger system size, one obtains a DTC phase for larger values of $\epsilon$ for a given value of $\sigma_{J'}$.  We thus conclude that the larger system size helps to further stabilize a DTC phase.

\subsection{Effects of leakage}
\begin{figure}[htb]
	\centering
	\includegraphics[width=\columnwidth]{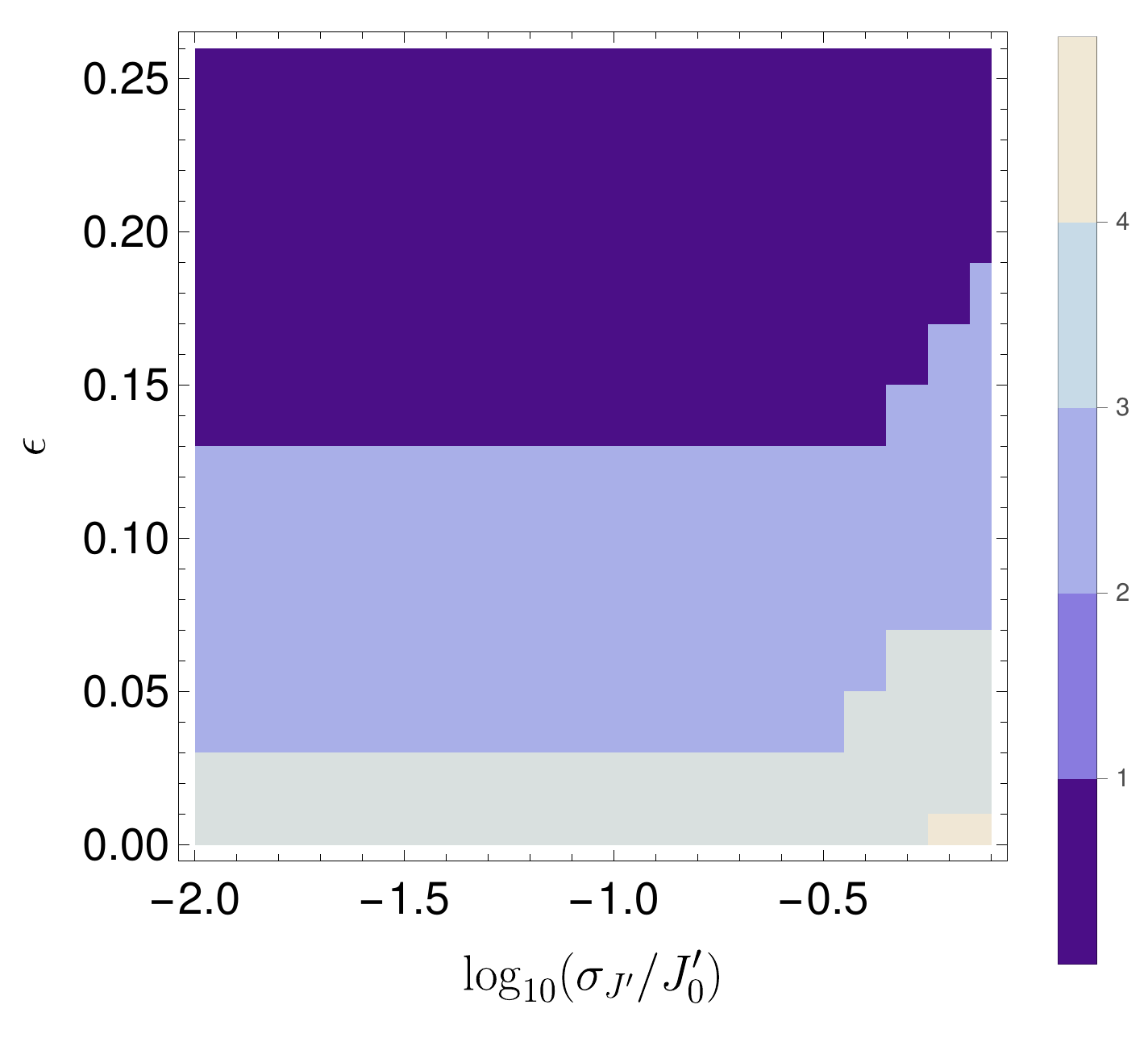}
	\caption{Plot of the number of states that display discrete time crystal (DTC) behavior as a function of $\epsilon$ and $\sigma_{J'}$ for six qubits with leakage and with a uniform magnetic field $B=0.5J'_0$ applied to the system.}
	\label{fig:PDFullPlotWL}
\end{figure}

\begin{figure}[htb]
	\centering
	\includegraphics[width=\columnwidth]{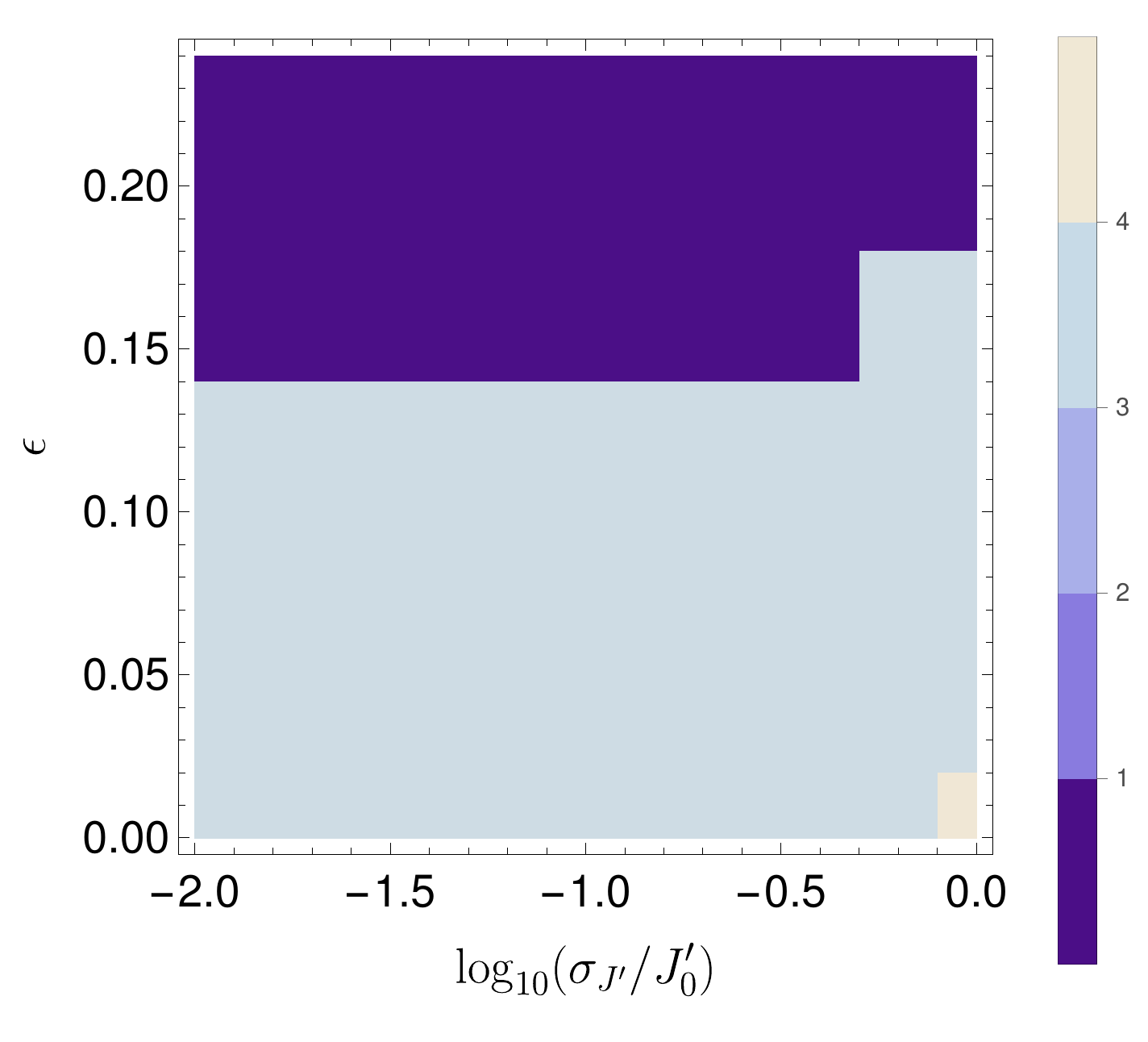}
	\caption{Plot of the number of states that display discrete time crystal (DTC) behavior as a function of $\epsilon$ and $\sigma_{J'}$ for eight qubits with leakage and with a uniform magnetic field $B=0.5J'_0$ applied to the system.}
	\label{fig:PDFullPlotWL_8Qubits}
\end{figure}

We now turn our attention to the effects of leakage, adding back in the leakage terms $H_L$ and the terms in $H_q$ and $H_\text{int}$ that we dropped in the no-leakage limit.  We begin with the case of a uniform overall applied magnetic field.  This is in fact the worst-case scenario for leakage.  While an overall magnetic field on one qubit will split off the two leakage states, as can be seen from Eq.~\eqref{eq:Hamiltonian_SQ_Full}, thus making it difficult for the qubit to enter these states on its own, the fact that the $z$ component of the total spin of the underlying Heisenberg exchange-coupled spin chain must be conserved means that qubits must ``leak'' in nearest-neighbor pairs and enter opposite leakage states (i.e., one must go into the $\ket{L_+}$ state, while the other must go into the $\ket{L_-}$ state).  This means that the energy cost for one qubit to enter a leakage state is ``paid'' by the other qubit entering the opposite leakage state, and thus the overall magnetic field actually has no effect on leakage.  In short, there is no energy difference between an overall system state in which two given nearest-neighbor qubits are both in the computational subspace and the same state, but with the two qubits in opposite leakage states.  We demonstrate this fact by repeating our previous calculations for six and eight qubits with the leakage terms added back in, and show our results for an applied overall magnetic field of $B=0.5J'_0$ in Figs.~\ref{fig:PDFullPlotWL} and \ref{fig:PDFullPlotWL_8Qubits}, respectively.  We see that leakage completely eliminates the DTC phase; at most, the system will be in a ``pre-thermal'' phase.

\begin{figure}[htb]
	\centering
	\includegraphics[width=\columnwidth]{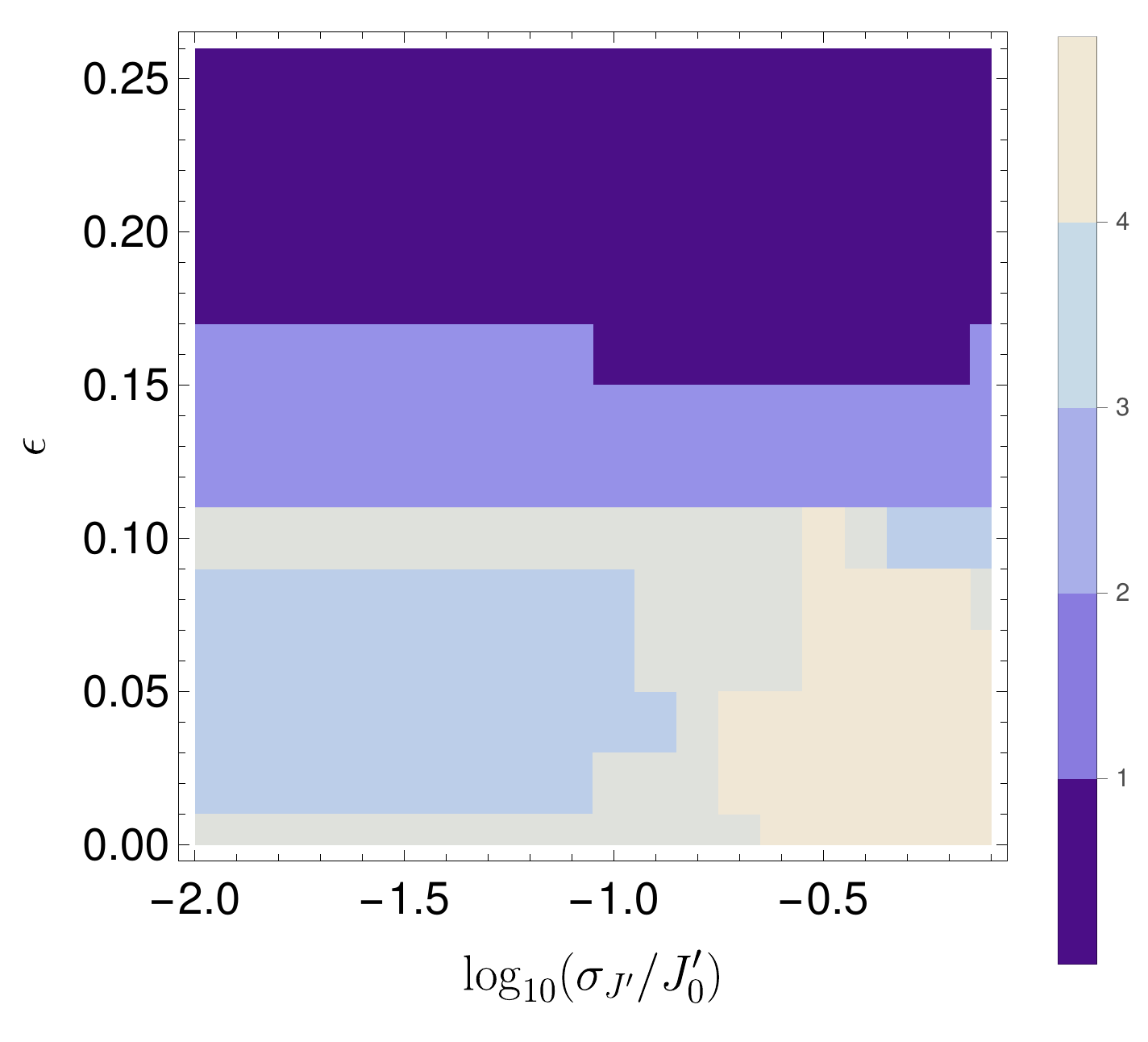}
	\caption{Plot of the number of states that display discrete time crystal (DTC) behavior as a function of $\epsilon$ and $\sigma_{J'}$ for six qubits with leakage and with an alternating magnetic field of magnitude $|B|=J'_0$ applied to the system.}
	\label{fig:PDFullPlotWL_SBF}
\end{figure}

\begin{figure}[htb]
	\centering
	\includegraphics[width=\columnwidth]{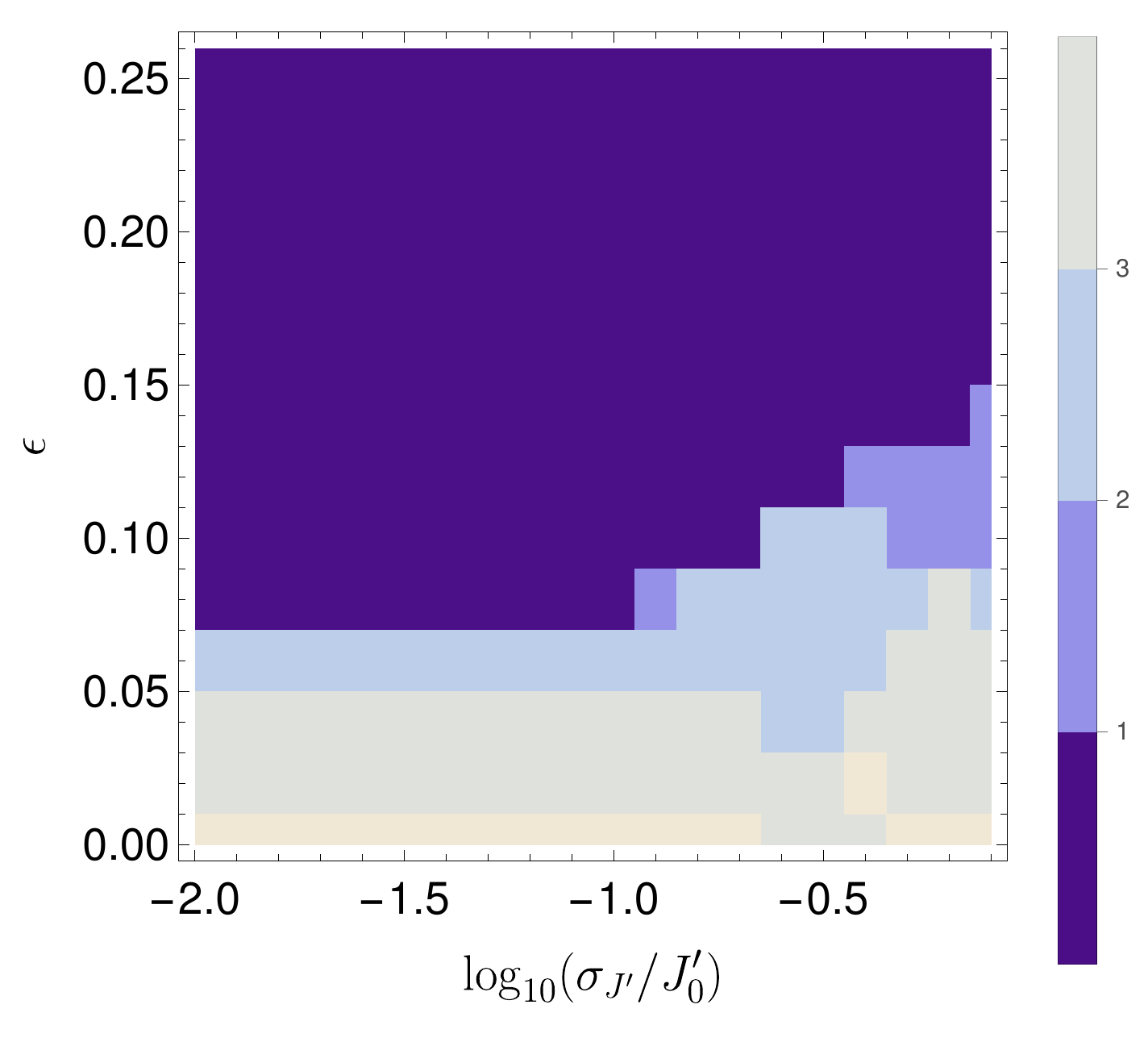}
	\caption{Plot of the number of states that display discrete time crystal (DTC) behavior as a function of $\epsilon$ and $\sigma_{J'}$ for six qubits with leakage and with an alternating magnetic field of magnitude $|B|=2J'_0$ applied to the system.}
	\label{fig:PDFullPlotWL_SBF_Med}
\end{figure}

\begin{figure}[htb]
\centering
\includegraphics[width=\columnwidth]{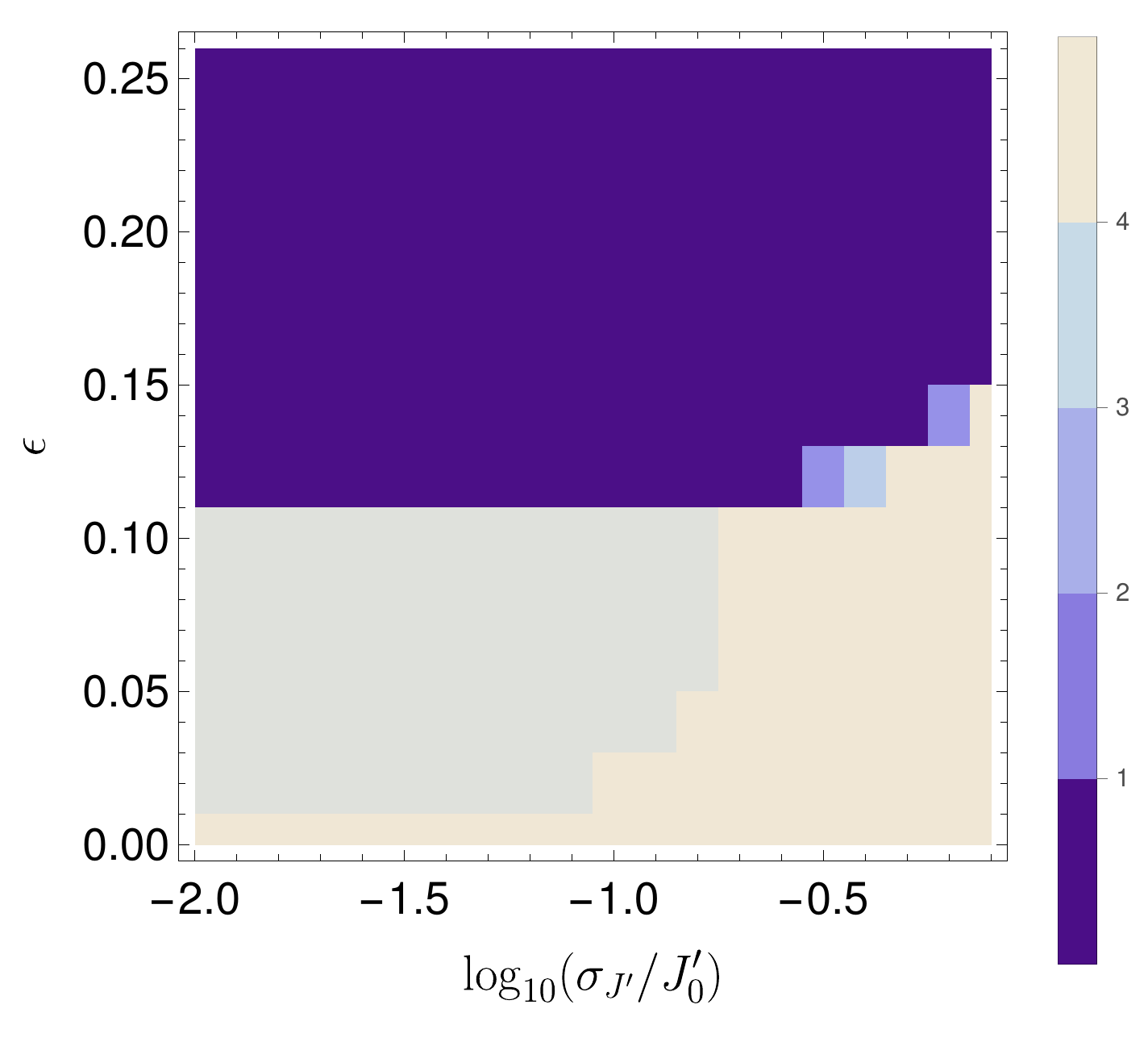}
\caption{Plot of the number of states that display discrete time crystal (DTC) behavior as a function of $\epsilon$ and $\sigma_{J'}$ for six qubits with leakage and with an alternating magnetic field of magnitude $|B|=10J'_0$ applied to the system.}
\label{fig:PDFullPlotWL_SBF_High}
\end{figure}

\begin{figure}[htb]
	\centering
	\includegraphics[width=\columnwidth]{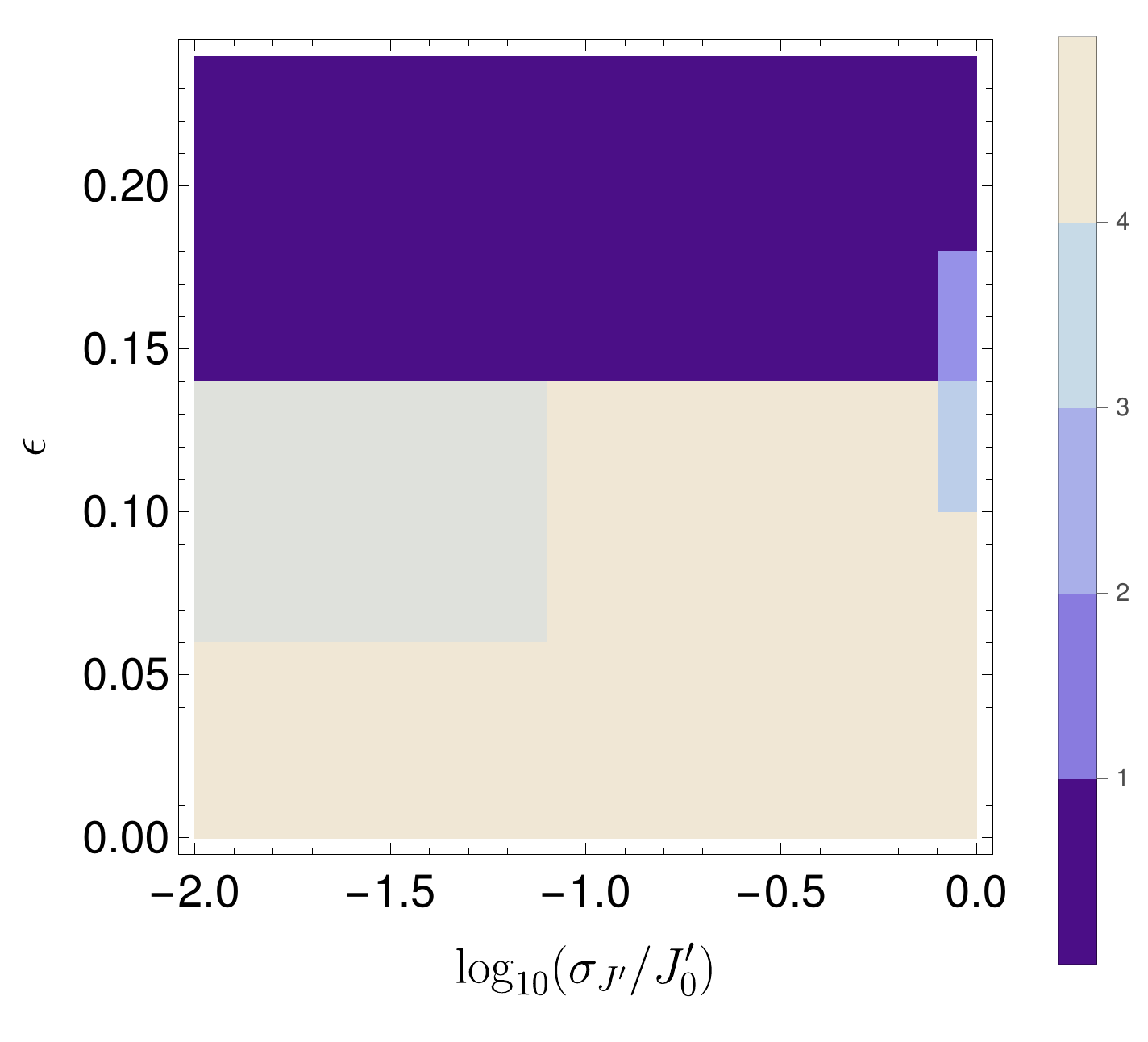}
	\caption{Plot of the number of states that display discrete time crystal (DTC) behavior as a function of $\epsilon$ and $\sigma_{J'}$ for eight qubits with leakage and with an alternating magnetic field of magnitude $|B|=10J'_0$ applied to the system.}
	\label{fig:PDFullPlotWL_SBF_High_8Qubits}
\end{figure}

The natural question that one may ask is whether there is any way to restore the DTC phase, even with leakage.  The answer is that it is indeed possible; configuring the overall magnetic field $B$ to alternate between qubits hinders leakage because, in this case, there is now an overall energy difference between the state with two given nearest-neighbor qubits in computational states and that with the qubits in leakage states, thus separating the leakage states of the overall system from the purely computational states.  We show results for an alternating magnetic field of magnitude $|B|=J'_0$ in Figs.~\ref{fig:PDFullPlotWL_SBF}.  We see that the DTC phase is restored, but for larger values of $\sigma_{J'}$ and lower values of $\epsilon$; we have thus mitigated, but not completely eliminated, the effects of leakage.
We also consider larger alternating magnetic fields.  We show results for $|B|=2J'_0$ for six qubits in Fig.~\ref{fig:PDFullPlotWL_SBF_Med} and for $|B|=10J'_0$ in Fig.~ \ref{fig:PDFullPlotWL_SBF_High}.
We note that, for $|B|=2J'_0$, the system no longer displays a DTC phase, but, for $|B|=10J'_0$, the results are hardly distinguishable from the no-leakage limit.  We see that, as expected, if the energy scale associated with the alternating magnetic field is much larger than that of the exchange couplings, then the alternating field ``freezes out'' the leakage states, thus effectively restoring the system to the no-leakage limit.  We find similar results for eight qubits in an alternating magnetic field of magnitude $|B|=10J'_0$; we show our results in Fig.~\ref{fig:PDFullPlotWL_SBF_High_8Qubits}.

\begin{figure}[htb]
	\centering
	\includegraphics[width=\columnwidth]{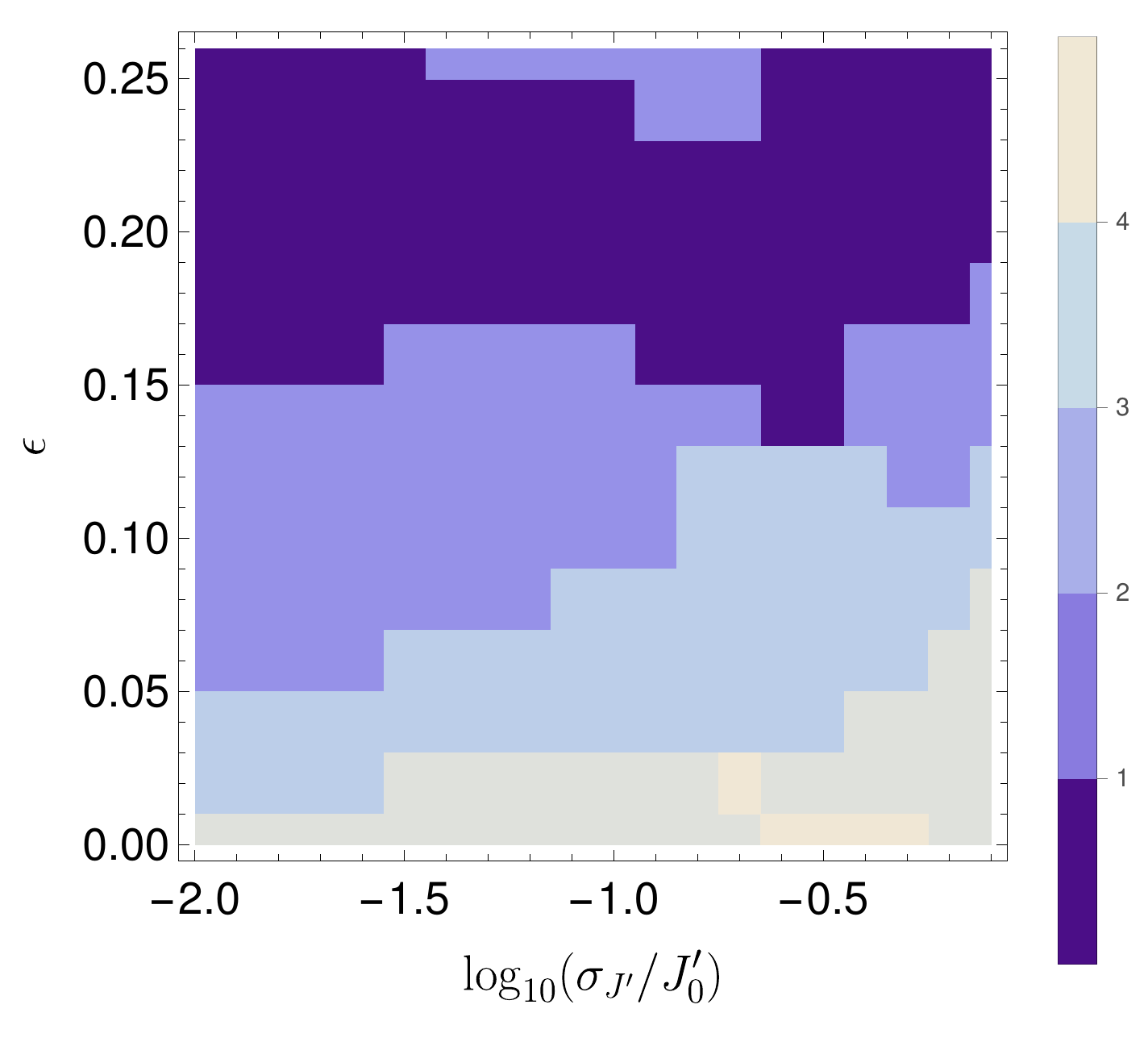}
	\caption{Plot of the number of states that display discrete time crystal (DTC) behavior as a function of $\epsilon$ and $\sigma_{J'}$ for six qubits with leakage and with a magnetic field of magnitude $|B|=J'_0$ in a ``two up, two down'' arrangement applied to the system.}
	\label{fig:PDFullPlotWL_2Up2Down}
\end{figure}

\begin{figure}[htb]
\centering
\includegraphics[width=\columnwidth]{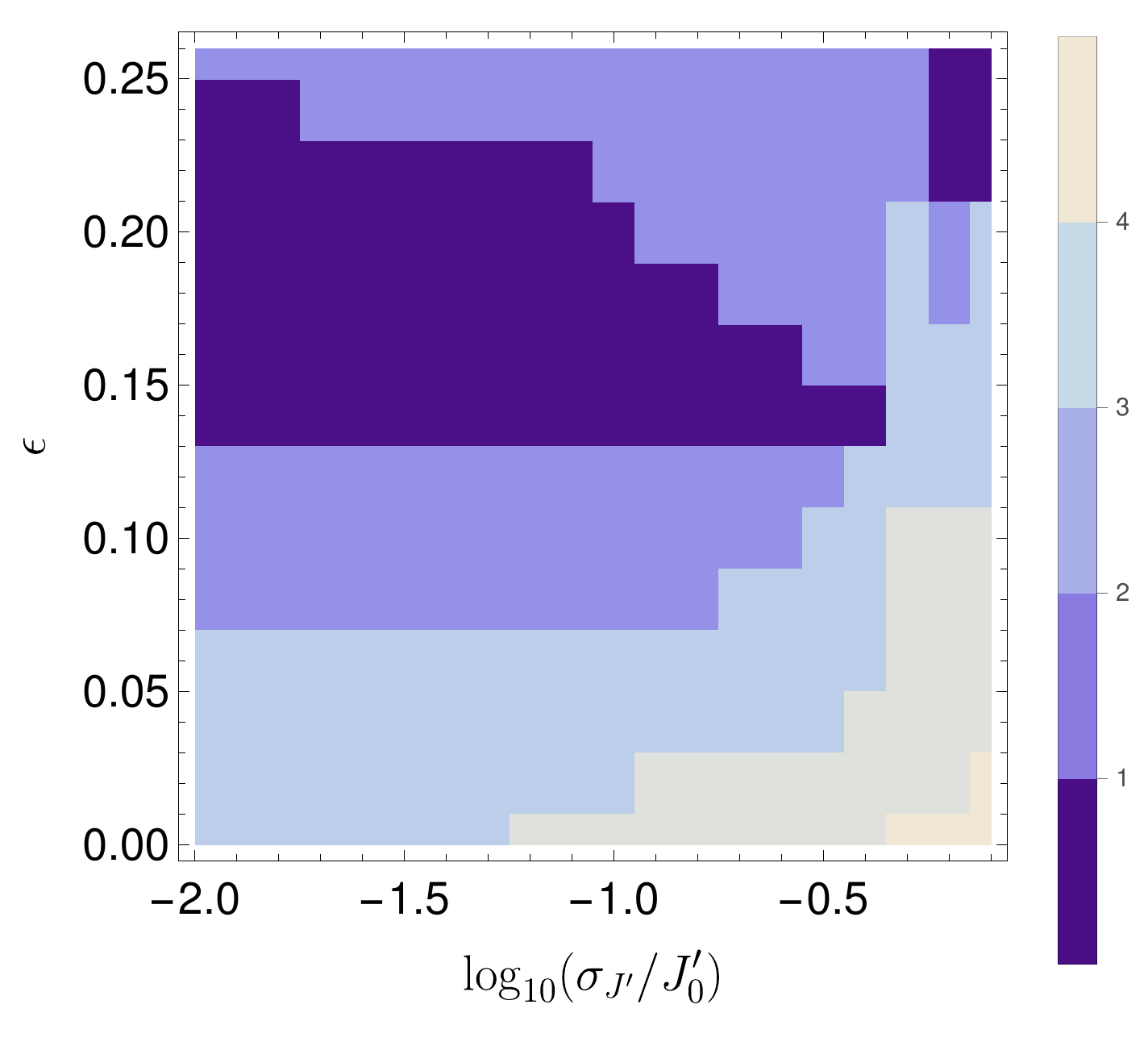}
\caption{Plot of the number of states that display discrete time crystal (DTC) behavior as a function of $\epsilon$ and $\sigma_{J'}$ for six qubits with leakage and with a magnetic field of magnitude $|B|=10J'_0$ in a ``two up, two down'' arrangement applied to the system.}
\label{fig:PDFullPlotWL_2Up2Down_High}
\end{figure}

\begin{figure}[htb]
\centering
\includegraphics[width=\columnwidth]{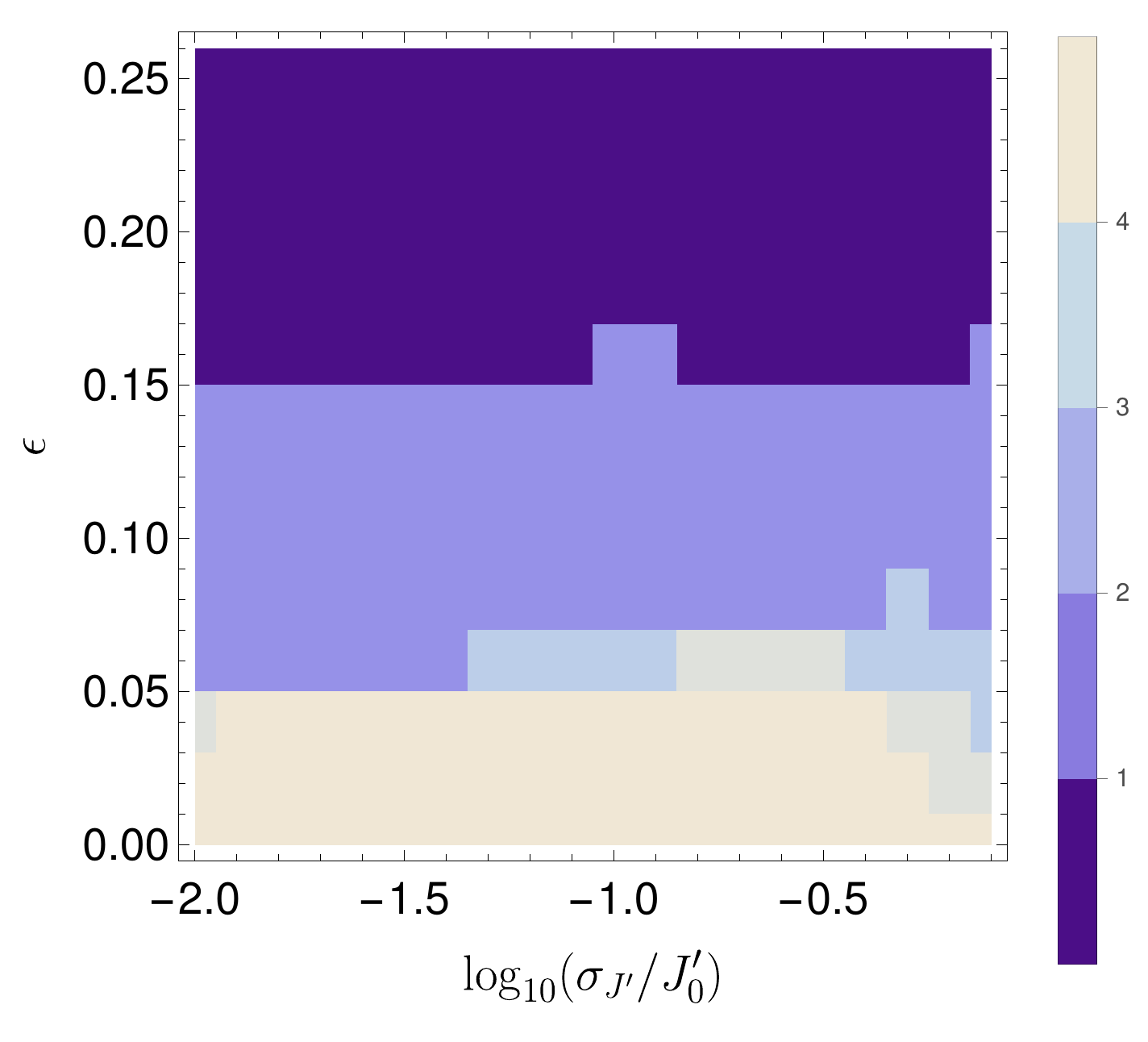}
\caption{Plot of the number of states that display discrete time crystal (DTC) behavior as a function of $\epsilon$ and $\sigma_{J'}$ for six qubits with leakage and with a magnetic field of magnitude $|B|=J'_0$ in a ``three up, three down'' arrangement applied to the system.}
\label{fig:PDFullPlotWL_3Up3Down}
\end{figure}

\begin{figure}[htb]
\centering
\includegraphics[width=\columnwidth]{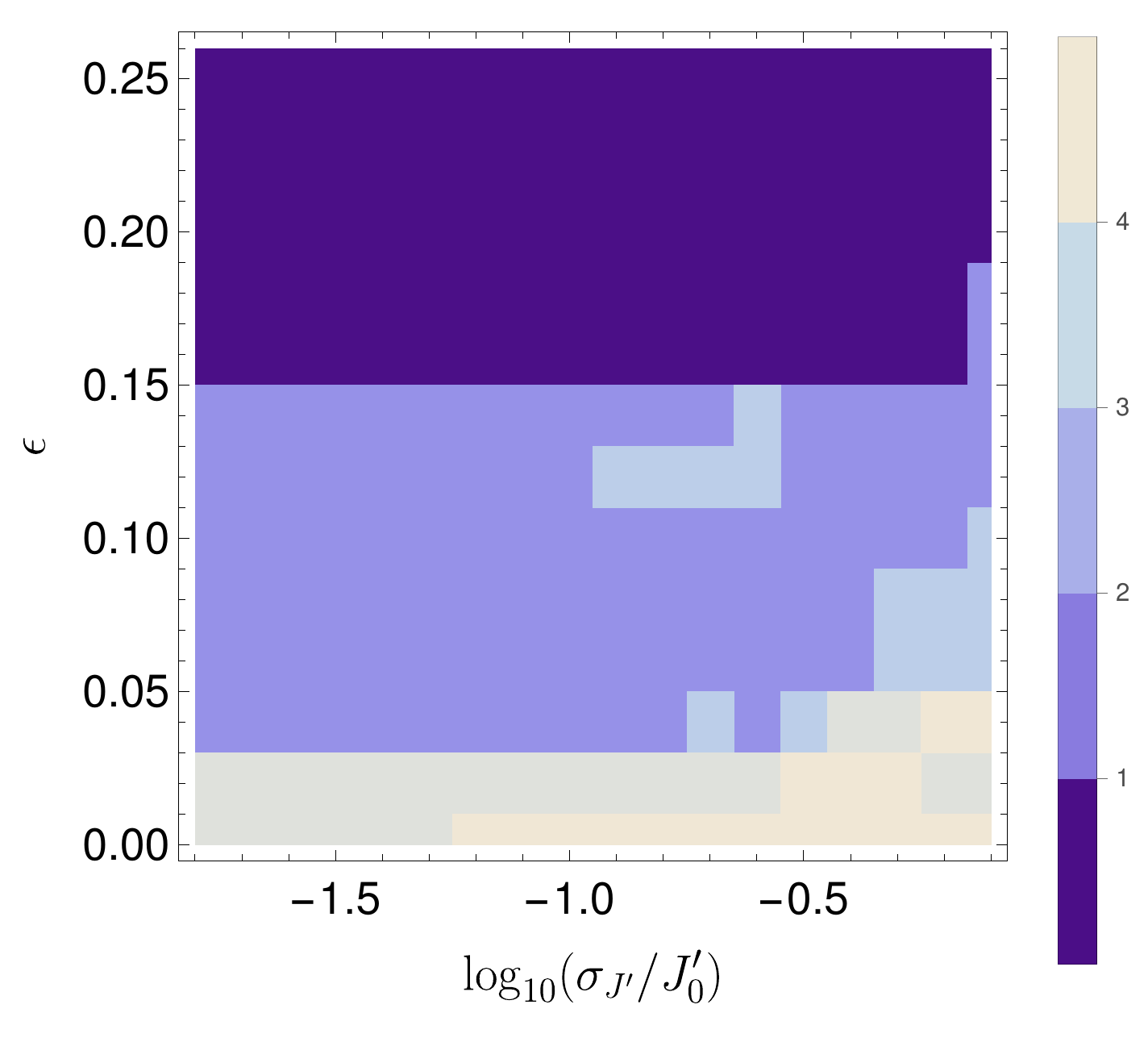}
\caption{Plot of the number of states that display discrete time crystal (DTC) behavior as a function of $\epsilon$ and $\sigma_{J'}$ for six qubits with leakage and with a magnetic field of magnitude $|B|=10J'_0$ in a ``three up, three down'' arrangement applied to the system.}
\label{fig:PDFullPlotWL_3Up3Down_High}
\end{figure}

\begin{figure}[htb]
\centering
\includegraphics[width=\columnwidth]{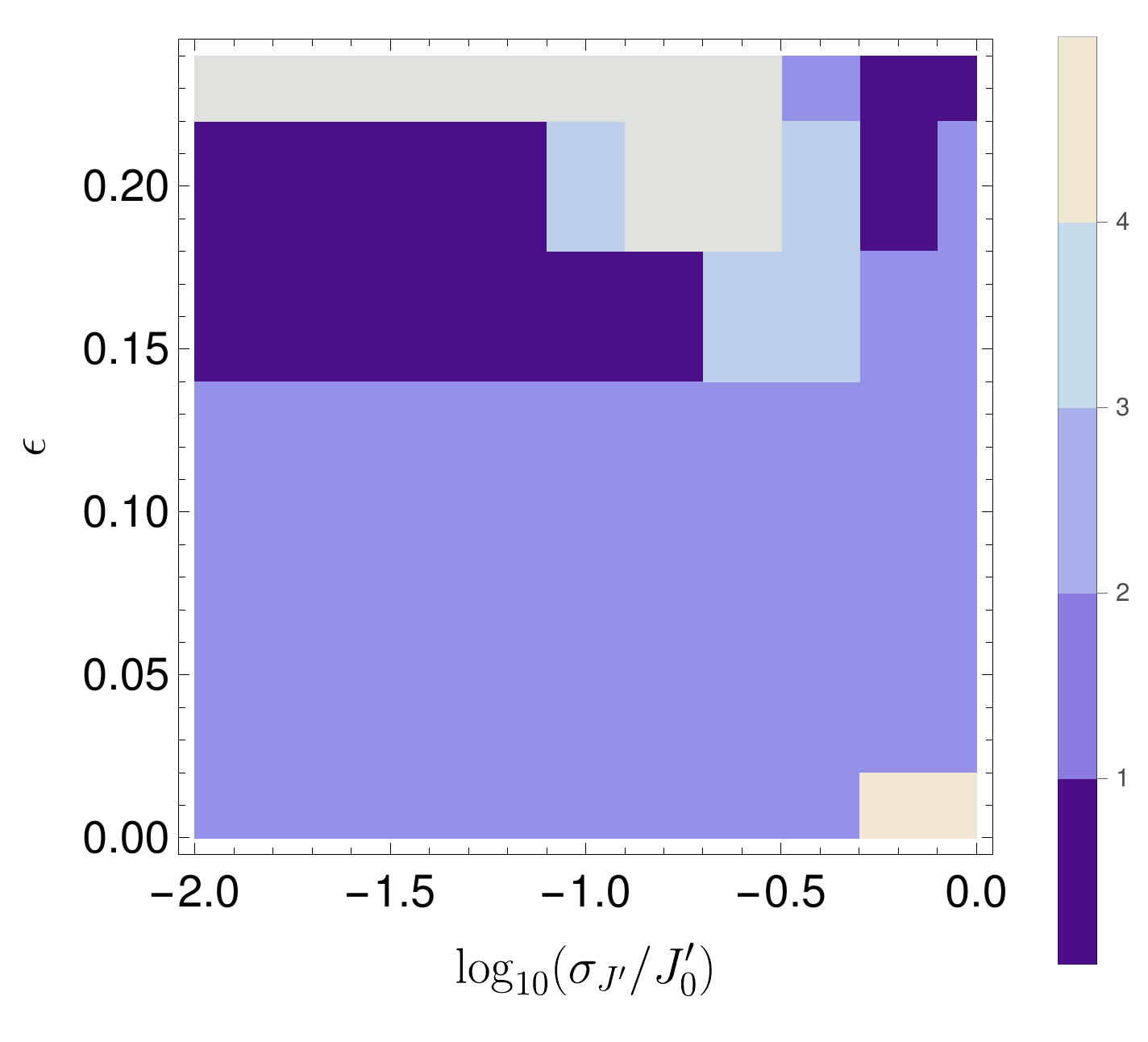}
\caption{Plot of the number of states that display discrete time crystal (DTC) behavior as a function of $\epsilon$ and $\sigma_{J'}$ for eight qubits with leakage and with a magnetic field of magnitude $|B|=10J'_0$ in a ``two up, two down'' arrangement applied to the system.}
\label{fig:PDFullPlotWL_2Up2Down_High_8Qubits}
\end{figure}

\begin{figure}[htb]
\centering
\includegraphics[width=\columnwidth]{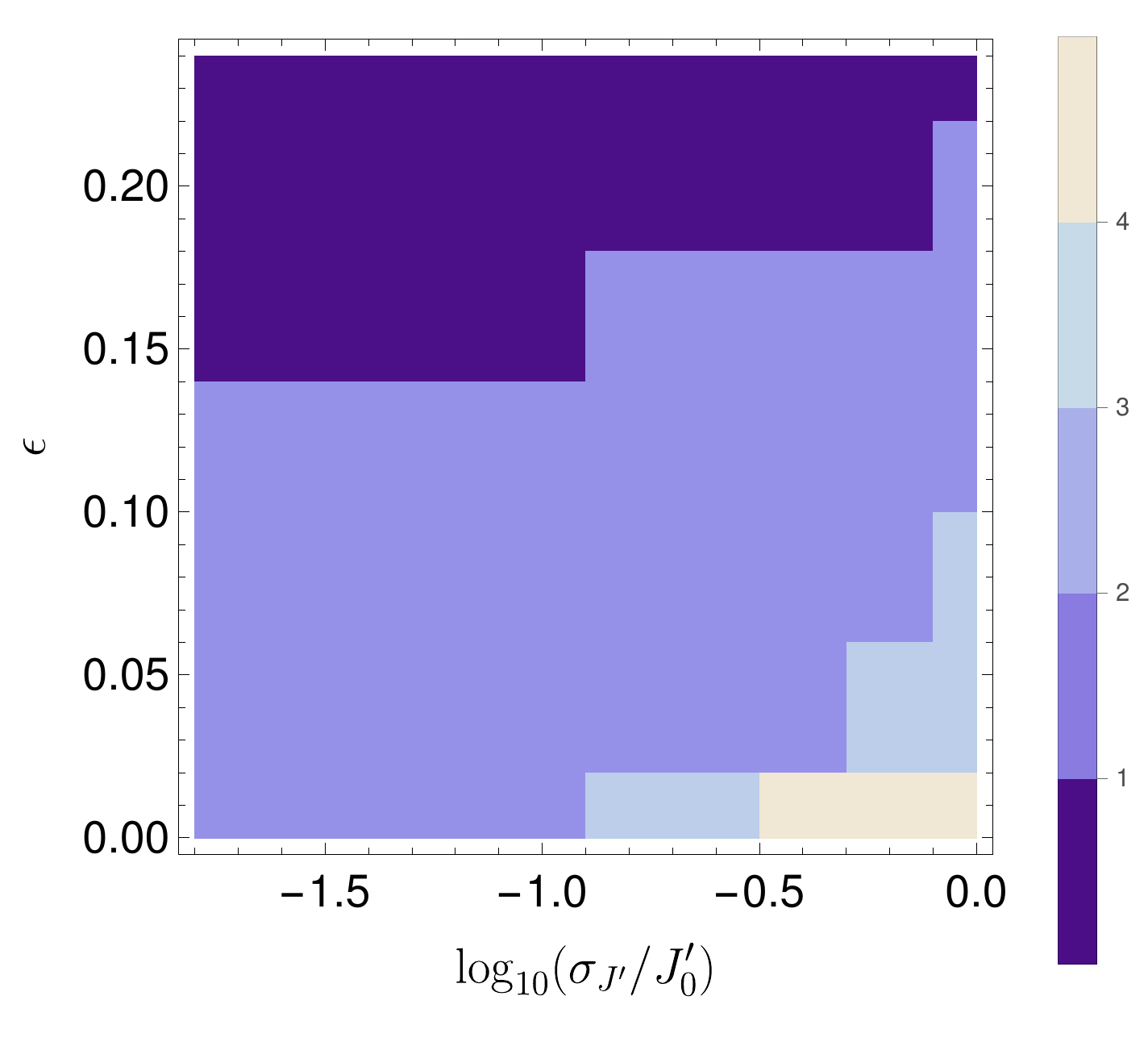}
\caption{Plot of the number of states that display discrete time crystal (DTC) behavior as a function of $\epsilon$ and $\sigma_{J'}$ for eight qubits with leakage and with a magnetic field of magnitude $|B|=10J'_0$ in a ``three up, three down'' arrangement applied to the system.}
\label{fig:PDFullPlotWL_3Up3Down_High_8Qubits}
\end{figure}

\begin{figure}[htb]
\centering
\includegraphics[width=\columnwidth]{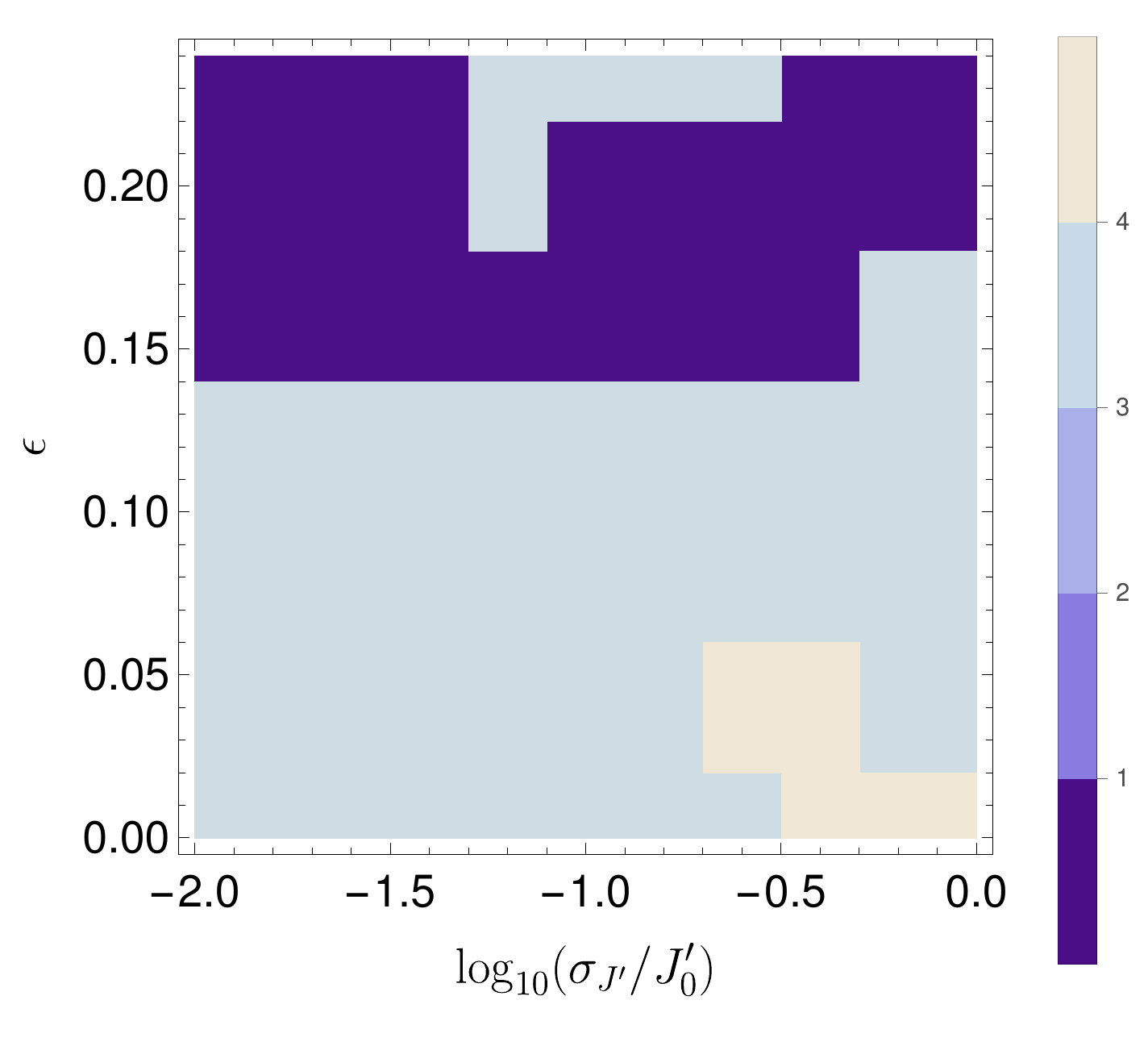}
\caption{Plot of the number of states that display discrete time crystal (DTC) behavior as a function of $\epsilon$ and $\sigma_{J'}$ for eight qubits with leakage and with a magnetic field of magnitude $|B|=10J'_0$ in a ``four up, four down'' arrangement applied to the system.}
\label{fig:PDFullPlotWL_4Up4Down_High_8Qubits}
\end{figure}

In addition to the alternating magnetic field arrangement, we also investigated two other arrangements---a ``two up, two down'' arrangement in which the first two qubits are subject to a field $+B$, then the next two are subject to a field $-B$, and so on, and a ``three up, three down'' arrangement in which the first three qubits are subject to a field $+B$, then the next three are subject to a field $-B$.  We plot our results for the ``two up, two down'' arrangement for $|B|=J'_0$ in Fig.~\ref{fig:PDFullPlotWL_2Up2Down} and for $|B|=10J'_0$ in Fig.~\ref{fig:PDFullPlotWL_2Up2Down_High}, and give the corresponding results for the ``three up, three down'' arrangement in Figs.~\ref{fig:PDFullPlotWL_3Up3Down} and \ref{fig:PDFullPlotWL_3Up3Down_High}, respectively.  We see that, for $|B|=10J'_0$, the ``two up, two down'' arrangement partially restores the DTC phase, but only for the largest $\sigma_{J'}$ value considered, while the ``three up, three down'' arrangement does not restore it at all.  We give the corresponding phase diagrams for all arrangements considered above for eight qubits and for $|B|=10J'_0$ in the nonuniform cases, plus results for a ``four up, four down'' arrangement (analogous to the ``two up, two down'' and ``three up, three down'' arrangements), in Figs.~
\ref{fig:PDFullPlotWL_2Up2Down_High_8Qubits}--\ref{fig:PDFullPlotWL_4Up4Down_High_8Qubits}.

We note that the results for $|B|=J'_0$ and $|B|=2J'_0$ are seemingly strange; they would imply that the ``three up, three down'' arrangement is better at restoring the DTC phase than either the ``two up, two down'' or alternating arrangements in some cases.  However, it is not surprising that we would see such behavior since the energy scale of the magnetic field is comparable to that of the interqubit exchange coupling, and thus the system's leakage states are not well separated from the purely computational states as they would be for $|B|=10J'_0$.

\section{Conclusion} \label{sec:Conclusion}
We investigated the possibility of realizing a discrete time crystal (DTC) phase in a chain of Heisenberg exchange-coupled quantum dot spins in an applied magnetic field being operated as a chain of singlet-triplet (ST) qubits.  Our main concern is leakage of the qubits out of the computational subspace.  Without the leakage terms present, the system would provide a realization of an Ising model with an applied magnetic field, which is a perfect system for realizing a DTC phase.  We considered systems with six and eight qubits and assumed the presence of quasistatic noise in both the exchange couplings and the magnetic field gradients, modeled as Gaussian distributions of the relevant terms in the Hamiltonian.

To determine whether or not the system is in a DTC phase, we considered how the system evolved, starting in four different initial conditions.  We then let the system evolve naturally for a time $T$, apply a pulse that performs a $(1-\epsilon)\pi$ rotation about the $z$ axis on all qubits, and repeat for $100$ iterations (Floquet cycles).  Here, $\epsilon$ represents an error in the rotation; ideally, the pulse would perform a $\pi$ rotation.  We determine the components of each qubit's state on the Bloch sphere as a function of the number of cycles and take the Fourier transform.  We fix the strength of the noise $\sigma_{\delta h}$ in the magnetic field gradients (i.e., the standard deviation of the Gaussian distribution) and vary the strength of the interqubit exchange coupling noise $\sigma_{J'}$ and the error in the qubit rotations $\epsilon$, determining how many of the initial conditions display DTC behavior.  In order for the system to be in a DTC phase for a given value of $\sigma_{J'}$, two criteria must be met: all four initial conditions must show a peak at a frequency $\omega=\pi/T$, corresponding to oscillations of period $2T$, for all qubits, and this peak must persist for $\epsilon>0$.

Based on our results, we constructed phase diagrams for a number of cases.  We begin with the no-leakage limit, in which all terms involving the leakage states are dropped, leaving only the effective Ising model terms.  We find a DTC phase in both cases for a large range of parameters for both six and eight qubits.  In fact, we find that the DTC phase exists over a larger parameter range in the eight-qubit case, showing that a larger system size helps to stabilize the DTC phase.  We then add back in the leakage terms.  We find that, if we only apply a uniform magnetic field to the system, then the leakage terms completely eliminate the DTC phase over the parameter range that we investigated.  Fortunately, we find that it is possible to mitigate the effects of the leakage terms by instead applying an alternating magnetic field; if the energy scale for the applied field is much larger than the interqubit exchange coupling, then we find that the resulting phase diagram is almost indistinguishable from the no-leakage case.  We also considered other arrangements of the magnetic fields that, rather than alternating at each qubit, alternate at every other qubit (``two up, two down''), every three qubits (``three up, three down''), and, in the eight-qubit case specifically, every four qubits (``four up, four down'').

We note that there is prior theoretical and experimental work on the realization of a DTC phase in an Ising model \cite{PhysRevB.99.035311,arxiv.2108.00942,science.abk0603,s41586-021-04257-w}, the model that we consider here, but our work focuses on a specific implementation of an Ising model, a chain of ST qubits, and on the issue of leakage out of the computational subspace that is specific to this implementation.  We also note that, in particular, the work of Ref.~\cite{PhysRevB.99.035311} also considers a method by which a Heisenberg spin chain may be converted into an effective Ising spin chain.  The method used there involves a special pulse sequence that converts the time evolution operator from that of a Heisenberg model to approximately that of an effective Ising model.  In contrast, we propose operating the Heisenberg spin chain as a chain of ST qubits, which, in the presence of an alternating magnetic field, realize an Ising model in the computational states.  We have shown that, even though leakage is a serious problem for the realization of a DTC in a chain of ST qubits, it is not insurmountable.  Given current experimental capabilities, it would be possible to realize a sufficiently strong alternating magnetic field to freeze-out the leakage states, thus allowing the realization of a DTC phase over a large range of parameters.  Our work thus suggests an experimental application of noisy intermediate-scale quantum (NISQ) systems, in this case systems consisting of $12$ or $16$ spins.  The observation of our predicted quantum dot DTC would bring spin qubits into the NISQ era of considerable current interest where qubits are used to achieve quantum tasks which are difficult (but not yet impossible) on classical computers.

\acknowledgments
This work is supported by the Laboratory for Physical Sciences.  The authors acknowledge the University of Maryland supercomputing resources (http://hpcc.umd.edu) made available for conducting the research reported in this paper.

\bibliography{DTC_STQubitChain}

\end{document}